\documentclass[%
 aip,
amsmath,amssymb,
reprint,%
]{revtex4-1}

\usepackage[bookmarks=false]{hyperref}
\DeclareMathAlphabet{\mathcal}{OMS}{cmsy}{m}{n}
\usepackage{graphicx}
\usepackage{dcolumn}
\usepackage{bm}

\usepackage[utf8]{inputenc}
\usepackage[T1]{fontenc}
\usepackage{mathptmx}
\usepackage{etoolbox}

\makeatletter
\def\@email#1#2{%
 \endgroup
 \patchcmd{\titleblock@produce}
  {\frontmatter@RRAPformat}
  {\frontmatter@RRAPformat{\produce@RRAP{*#1\href{mailto:#2}{#2}}}\frontmatter@RRAPformat}
  {}{}
}%
\makeatother

\renewcommand{\v}[1]{\ensuremath{\bm{#1}}} 
 
\newcommand{\uv}[1]{\ensuremath{\bm{#1}}} 
\newcommand{\pd}[2]{\frac{\partial #1}{\partial #2}} 
 
\let\baraccent=\= 
\renewcommand{\=}[1]{\stackrel{#1}{=}} 

\newcommand{\bhat}{\uv{b}}

\newcommand{\upari}{u_{\parallel i}}
\newcommand{\upare}{u_{\parallel e}}

\newcommand{\jac}{\mathcal{J}}

\newcommand{\vpar}{v_{\parallel}}

\newcommand{\gkyl}{\texttt{Gkeyll}}

\newcommand{\ExB}{$E\times B$}

\begin{document}




\title{Kinetic modeling of neutral transport for a continuum gyrokinetic code}

\author{T.~N.~Bernard}
    \email{bernardt@fusion.gat.com}
\affiliation{General Atomics, San Diego, CA 92186, USA}
\author{F.~D.~Halpern}
\affiliation{General Atomics, San Diego, CA 92186, USA}
\author{M. Francisquez}
\affiliation{%
Princeton Plasma Physics Laboratory, Princeton, NJ 08543, USA
}%
\author{N.~R.~Mandell}
\affiliation{MIT Plasma Science and Fusion Center, Cambridge, MA 02139, USA
}%
\author{J. Juno}
\affiliation{%
Princeton Plasma Physics Laboratory, Princeton, NJ 08543, USA
}%
\affiliation{Department of Physics and Astronomy, University of Iowa, Iowa City, IA 52242, USA}
\author{G.~W.~Hammett}
\affiliation{%
Princeton Plasma Physics Laboratory, Princeton, NJ 08543, USA
}%
\author{A.~Hakim}
\affiliation{%
Princeton Plasma Physics Laboratory, Princeton, NJ 08543, USA
}%
\author{G. Wilkie}
\affiliation{%
Princeton Plasma Physics Laboratory, Princeton, NJ 08543, USA
}%
\author{J. Guterl}
\affiliation{General Atomics, San Diego, CA 92186, USA}


\date{\today}

\begin{abstract}
We present the first-of-its-kind coupling of a continuum full-$f$ gyrokinetic turbulence model with a 6D continuum model for kinetic neutrals, carried out using the \gkyl\ code. Our objective is to improve the first-principles understanding of the role of neutrals in plasma fueling, detachment, and their interaction with edge plasma profiles and turbulence statistics. Our model includes only atomic hydrogen and incorporates electron-impact ionization, charge exchange, and wall recycling. These features have been successfully verified with analytical predictions and benchmarked with the DEGAS2 Monte Carlo neutral code. We carry out simulations for a scrape-off layer (SOL) with simplified geometry and NSTX parameters. We compare these results to a baseline simulation without neutrals and find that neutral interactions reduce the normalized density fluctuation levels and associated skewness and kurtosis, while increasing auto-correlation times. A flatter density profile is also observed, similar to the SOL density shoulder formation in experimental scenarios with high fueling. 
\end{abstract}


\maketitle 

\section{Introduction}
Neutral particles are present in the edge plasma of magnetic fusion devices due to plasma interactions with wall and divertor materials, as well as beam injection and neutral gas puffs. Atomic and molecular neutrals interact with plasma particles via electron-impact ionization, charge exchange, radiative recombination, etc. Neutrals can penetrate into the core and affect plasma dynamics through these collisional processes. They play a role in plasma fueling via recycling, whereby ions impinging on the wall are re-emitted as neutrals, which then become ionized. Furthermore, ion--neutral friction is necessary to reduce the heat flux to the divertor in detached scenarios.\cite{stangeby2000plasma} The effect of neutrals on confinement has been investigated experimentally\cite{Boivin2000,Groebner2002,Moser2020,Mordijck2020} and accurate theoretical and numerical models are necessary to thoroughly interpret those results.

These experimental studies have generated interest in coupling models of neutral transport to first-principles plasma turbulence codes. Couplings with fluid neutral models have effectively modeled some scrape-off layer (SOL) conditions,\cite{Thrysoe2018,Zholobenko2021} but are rigorous only where the neutral mean free paths are shorter than the characteristic length scale. Comparisons between a fluid neutral code and a kinetic model were improved recently with a more accurate numerical treatment of geometry.\cite{dekeyser2019implementation} Based on typical SOL conditions, neutral mean free paths can vary from several centimeters to a meter or more.\cite{stangeby2000plasma} One characteristic length scale is the size of coherent turbulent structures in the plasma, or blobs, which are on the order of centimeters.\cite{dippolito2011} Kinetic models for neutrals are necessary to accurately capture both long and short mean free path neutrals. Because of this, much neutral modeling has been carried out by kinetic Monte Carlo (MC) codes,\cite{reiter2002,stotler1994,cupini1983nimbus} and these models have been coupled to fluid turbulence codes like TOKAM2D~\cite{Marandet2013} and TOKAM3X.\cite{Galassi2017,Fan2019} SOLPS has been coupled to the six-field two-fluid turbulence code BOUT++~\cite{zhang2019}, and the XGC framework incorporates a model of MC neutrals.\cite{Stotler2013,Stotler_2017,Ku2018,Hager2019}
MC codes are subject to statistical noise, which can interfere with the accuracy and convergence of Eulerian codes\cite{JOSEPH2017813} when coupled to them. Thus, various continuum kinetic models for neutrals have also been developed and coupled to first-principles turbulence codes such as GBS\cite{Wersal_2015,Wersal2017,Wersal2017b,mandell2020electromagnetic,Mancini2021} and nSOLT.\cite{Russell2019,Russell2021}

Gyrokinetic models, which have long been used to model core plasma turbulence, have been more recently adapted to the edge and SOL to capture important kinetic effects, such as parallel transport, trapped particles, nonlinear wave--particle interactions, etc., that drift-reduced fluid and gyrofluid codes may model less accurately. \cite{Pan2016,Chang2017,shi2019full,boesl2019gyrokinetic,Dorf2020,mandell2020electromagnetic,Michels2021} 
Including neutral interactions further improves the predictive capabilities of these models. Thus, in this work we present a continuum kinetic model of neutral transport coupled to a continuum gyrokinetic solver within the computational plasma physics framework \gkyl,\cite{gkeyllWeb} which has previously been used to model plasma turbulence on open field lines.\cite{shi2017gyrokinetic,shi2019full,bernard2019,mandell2020electromagnetic,Hakim2020,Bernard2020} The collisionless Vlasov solver in \gkyl \cite{Juno2018,HakimJuno:2020} facilitated the implementation of this model. A continuum kinetic neutral model avoids the statistical noise issues that are associated with MC codes and the shortcomings of fluid neutral transport models. The model only includes atomic neutrals and the electron-impact ionization and charge exchange processes. We model wall recycling as a boundary condition in the direction parallel to the magnetic field. Recombination will eventually be included, when high-recycling divertor scenarios are considered. For plasmas considered in this work, where $T_e \geq 10$ eV, this interaction is negligible. Other important reactions to include in future work are radiation, molecular processes, ion--neutral elastic collisions, and neutral--neutral collisions. 

The results presented here comprise the first-of-its-kind coupling of a continuum gyrokinetic solver to a continuum kinetic model for neutral transport. We have verified the neutral model against analytic theory in low dimensional tests with static plasma species. Benchmark tests, also in low dimensions with static plasma species, show good agreement with the DEGAS2 MC neutral code. 

We have demonstrated proof-of-concept in a high-dimensional simulation in three physical space dimensions, and two and three velocity space dimensions for the plasma species and neutral species, respectively (3X2V+3X3V). We model the National Spherical Torus Experiment (NSTX) SOL with simplified helical geometry and dynamic plasma species, similar to previous NSTX simulations with Gkeyll.\cite{shi2019full,mandell2020electromagnetic,Mandell2021thesis} We compare the results to a simulation without neutrals and with the same midplane plasma source. As expected, the inclusion of neutrals results in an increase in the steady state density due to ionization sourcing. A flattening of the density profile is also observed, and the ratio of the radial to parallel flux increases. These results support the conclusion in Ref.~\citenum{Mancini2021} that a flatter density profile can form in parameter regimes where the radial flux becomes more efficient at transporting density relative to the parallel flux. The electron and ion temperatures decrease due to energy loss occurring through the ionization process, transfer of energy to neutrals via charge exchange, and energy exchange via electron--ion collisions. In this simulation, we find that neutrals do not have a large impact on the parallel heat flux width. Power balance calculations demonstrate that energy is conserved to within 5\% of the input power.

The article is organized as follows. Section \ref{sec:model-eqns} describes the model equations used in \gkyl\ and the simplified model for neutral transport that has been implemented. Section~\ref{sec:1d-sol} contains results from SOL tests in one spatial dimension with static plasma species, including comparisons with analytic predictions and benchmarks with DEGAS2. Section~\ref{sec:nstx} contains results from high-dimensional simulations of a SOL with NSTX parameters and an analysis of the effect of neutrals. Section~\ref{sec:concl} summarizes our findings. The simulations that produce the results contained in this paper can
be reproduced with the input files made available online (see Appendix \ref{apx:get-gkyl}).

\section{Model equations for coupled gyrokinetic plasma and kinetic neutral species}\label{sec:model-eqns}
In this section we describe the equations that \gkyl\ solves, including conduction sheath and wall recycling boundary conditions. The simplified models for electron-impact ionization and charge exchange that have been implemented are also explained in detail here.
\subsection{Plasma species}
To model plasma transport on open-field lines in fusion devices, 
\gkyl\ evolves the full-$f$ gyrokinetic distribution function $f_s(\bm{x},\vpar,\mu,t)$ in the long-wavelength (or drift-kinetic) limit using the following equation:
\begin{eqnarray}
    \pd{\jac_s f_s}{t} + \nabla \cdot (\jac_s \{\bm{R},H\} f_s) + \pd{}{\vpar} (\jac_s \{v_\parallel, H\} f_s ) \nonumber \\ =\jac_s C[f_s] + \jac_s S_s.      \label{eq:gk} 
\end{eqnarray}
While \gkyl\ can handle electromagnetic fluctuations,\cite{mandell2020electromagnetic} for simplicity here we will consider only electrostatic fluctuations. The collision operator $C[f_s]$ can include Coulomb collisions, elastic collisions, and inelastic neutral interactions, and $S_s$ is a source term. Coulomb collisions are modeled using a Dougherty collision operator \cite{Francisquez2020lbo} and collisional interactions with neutrals are described in the following subsection. The Jacobian is $\jac = B^*_\parallel = \bm{b} \cdot \bm{B}_\parallel^*$, where $\bm{B}_\parallel^* = \bm{B} +(Bv_\parallel/\Omega_s)\nabla \times \bm{b}$. The Poisson brackets are defined by  
\begin{eqnarray}
    \{F,G\} = \frac{\bm{B}^*}{m_s B_\parallel^*} \cdot \left( \nabla F \frac{\partial G}{\partial v_\parallel} - \frac{\partial F}{\partial v_\parallel} \nabla G \right) \nonumber \\
- \frac{1}{q_s B_\parallel^*} \bm{b} \cdot \nabla F \times \nabla G.
\label{eq:pb}
\end{eqnarray}
In the long-wavelength limit, the Hamiltonian is $H_s = \frac{1}{2}mv_\parallel^2 + \mu B + q_s \phi$, and the system is closed by the gyrokinetic Poisson equation
\begin{equation}
     -\nabla \cdot \left( \frac{n_{i0}^g q_i^2 \rho_{\mathrm{s}0}^2}{T_{e0}} \nabla_\perp \phi \right) = \sigma_g
 = q_i n_i^g(\bm{R},t) - e n_e(\bm{R},t),
 \label{eq:poisson}
\end{equation}
which uses a linearized form for the polarization term, similar to the Boussinesq approximation made in Braginksii fluid codes.

A non-orthogonal field-line-following coordinate system is used for the configuration space grid, where $z$ is the distance along the field, $x$ is the radial coordinate, and $y$ is the binormal coordinate. Plasma and neutral species share the same configuration space grid. For the plasma species, Dirichlet boundary conditions with $\phi = 0$ are used in the radial direction and periodic boundary conditions are used in the binormal direction. Conducting-sheath boundary conditions are applied to $f$ in the $z$ direction, by which one plasma species is partially reflected and the other is completely absorbed.\cite{shi2017gyrokinetic} The gyrokinetic Poisson equation \ref{eq:poisson} is solved across the entire spatial domain, and the values at $z_{\min}$ and $z_{\max}$ (or $x_{\min}$ and $x_{\max}$ in Sec.~\ref{sec:1d-sol}) give the electrostatic potential at the sheath entrance. For simulations presented in subsequent sections, the sheath potential is positive, and it thus determines a cutoff velocity for the electron species, below which they are reflected. Ions free stream out of the domain at the parallel boundaries and are recycled as neutrals when coupled to the Vlasov neutral solver. 

\subsection{Neutral species}
The neutral particle distribution function $f_n(\bm{x},\bm{v},t)$ is evolved using the Vlasov equation
\begin{equation}
    \pd{f_n}{t} + \bm{v} \cdot \nabla f_n = C[f_n] + S_n, \label{eq:vlasov}
\end{equation}
where $C[f_n]$ can represent elastic collisions (not included in present results), which are modeled by a Bhanagar-Gross-Krook (BGK) operator, and also inelastic neutral interactions such as ionization and charge exchange. $S_n$ is a volumetric source term. As previously mentioned, the neutral distribution function is evolved on the same configuration space grid as the plasma species. Though the neutrals are evolved in the field-line-following configuration space, they are not affected by the magnetic field. The velocity space grid includes three velocity space coordinates which are orthogonal ($v_R,v_Z,v_\varphi$), where subscripts ($R,Z,\varphi$) correspond to the coordinates in a cylindrical system. Simulations in this work assume a simplified helical magnetic geometry, and the details of the neutral dynamical equation in this geometry are contained in Appendix \ref{apx:geo}.

A model for wall recycling has been implemented at the boundaries where field lines terminate at the endplate. These boundary conditions provide a source of neutrals from the targets that depends on the flux of incident ions. Consider a simulation with one configuration space dimension, parallel to the magnetic field, and three velocity space dimensions (1X3V). Coordinate $z$ is parallel to the magnetic field, and $v_z$ is the velocity for neutrals corresponding to that direction. We define the neutral distribution function in the ghost cell at the lower ($z_{\min}$) boundary as
\begin{eqnarray}
    f_n(v_x,v_y,v_z,z=z_{ghost}) = \alpha_b f_n(v_x,v_y,-v_z, z_{\min}) \nonumber \\ + C_{rec} f_{M,rec}(T= T_{n,rec}) \label{eq:wr1},
\end{eqnarray}
where $\alpha_b$ is the reflection coefficient. The unit-density Maxwellian function for recycled neutrals $f_{M,rec}$ is defined by a zero mean flow and a user-specified temperature of $T_{n,rec} = 10$ eV for simulations presented here. (Typical Franck-Condon temperatures of 2 eV require a finer velocity space grid and are computationally expensive. Non-uniform velocity space grids would alleviate this issue and are being explored for future simulations.) The Maxwellian is scaled such that the magnitude of the flux of incoming neutrals is equal to the magnitude of the flux of incident ions. 
\begin{eqnarray}
    C_{rec} \int_{0}^{v_{z,\max}} \int_{v_{y,\min}}^{v_{y,\max}} \int_{v_{x,\min}}^{v_{x,\max}} dv_x dv_y dv_z  \; v_z \, f_{M,r} \nonumber \\ 
    \doteq \alpha_{rec} \frac{2\pi}{m}\int_{0}^{\mu_{\max}}\int_{v_{\parallel,i,\min}}^0 d\vpar d\mu \;  v_{\parallel,i} \jac f_i(z=z_{\min}) \label{eq:wr2} \nonumber \\
    C_{rec} = \frac{\alpha_{rec} M_{1z-,i}}{M_{1z+,n}},
\end{eqnarray}
where $\alpha_{rec}$ is the user-specified recycling fraction, $\jac$ is the Jacobian, $\doteq$ denotes `weak equality' according to the modal discontinuous Galerkin (DG) scheme used in \gkyl\cite{hakim2020lbo}, and + or - subscripts refer to partial moments calculated over $v_z > 0$ or $v_z < 0$, respectively. The scaling coefficient $C_{rec}$ is calculated dynamically in the simulation. An angular dependence in the recycling boundary conditions will be included in future work. The recycling boundary conditions in \gkyl\ are not currently used to model recycling at the outer wall due to the zero-flux, conducting wall boundary condition used for the plasma species at the radial boundaries.

\subsection{Plasma-neutral interactions}
Here we present the details of the simplified  neutral interaction models that have been implemented in \gkyl. Electron-impact ionization, described by the process $e^{-} + n \rightarrow i^{+} + 2e^{-} - E_{iz}$, can be rigorously modeled by a collision term containing an integral over velocity space:
\begin{eqnarray}
    C^{iz}_e = C^{iz}_i = -C^{iz}_n \nonumber \\ 
    = f_n(\bm{v}) \int f_e(\bm{v}') \sigma_{iz}(|\bm{v} - \bm{v}'|) |\bm{v} - \bm{v}'| \, d\bm{v}'. \label{eq:iz-full}
\end{eqnarray}
Assuming that the electron thermal speed is much greater than the neutral thermal speed, $|\bm{v}'| \gg |\bm{v}|$, this simplifies to 
\begin{eqnarray}
    \int f_e(\bm{v}') \sigma_{iz}(|\bm{v} - \bm{v}'|) |\bm{v} - \bm{v}'| \, d\bm{v}' \nonumber \\ 
    \approx \int f_e(\bm{v}') \sigma_{iz}(v') v'\, d\bm{v}' = n_e \langle \sigma_{iz} v_e \rangle, \label{eq:iz-reduc}
\end{eqnarray}
where $\langle \sigma_{iz} v_e \rangle$ is the ionization rate parameter, whose value can be obtained from atomic databases for various elements. In \gkyl, the ionization reaction rate is approximated with a fitting formula \cite{Voronov1997}
\begin{equation}
    \langle \sigma_{iz} v_e \rangle = A  \frac{1 + P \, (E_{iz}/T_e)^{1/2}}{X + E_{iz}/T_e} \left(\frac{E_{iz}}{T_e}\right)^K e^{-E_{iz}/T_e} \, \times 10^{-6} {\rm m}^{3}/{\rm s}, \label{eq:vor}
\end{equation}where constants $A$, $P$, $X$ and $K$ are tabulated for elements up to $Z=28$. $E_{iz}$ is the ionization energy and is 13.6 eV for simulations presented in subsequent sections. This model has been used previously in \gkyl\ as part of Vlasov-Maxwell simulations of a plasma sheath.\cite{cagas2017continuum,Cagas2018} It was included as a source term in the plasma species equations to help the sheath relax to a steady state, and lower-energy electrons that resulted from the process were not accounted for. 

We model the ionization collision term for the electron species based on the neutral model in Ref.~\citenum{Wersal_2015}
\begin{equation}
    C^{iz}_e= n_n \langle \sigma_{iz} v_e \rangle \left[2f_{M,iz}(n_e,\bm{u}_n,v^2_{t,iz}) - f_e \right], \label{eq:iz-elc}
\end{equation}
where $\bm{u}_n$ is the neutral fluid velocity and $v^2_{t,iz}$ is the thermal velocity of the resulting lower-energy electrons, given by $v^2_{t,iz} = v^2_{t,e}/2 - E_{iz}/(3m_e)$. Finally, the ion and neutral collisional ionization terms are given by
\begin{eqnarray}
    C^{iz}_i &= n_e f_n \langle \sigma v_e \rangle, \label{eq:iz-ion}\\
    C^{iz}_n &= -n_e f_n \langle \sigma v_e \rangle. \label{eq:iz-neut}
\end{eqnarray}
These terms depend on moments of the electron and neutral distribution function such as $n_e$, $v_{t,e}^2$, and $\v{u}_n$, which are calculated at every timestep. The electron temperature appearing in Eq.~9 is obtained from the thermal speed, $v_{t,e}^2 = T_e/m_e$. 

The charge exchange process can be described by the collisional term  
\begin{eqnarray}
    C^{cx}_i = -C^{cx}_n = f_n(\bm{v}) \int f_i(\bm{v}') \sigma_{cx}(|\bm{v} - \bm{v}'|) |\bm{v} - \bm{v}'| \, d\bm{v}'  \nonumber \\ 
    - f_i(\bm{v}) \int f_n(\bm{v}') \sigma_{cx}(|\bm{v} - \bm{v}'|) |\bm{v} - \bm{v}'| \, d\bm{v}', \label{eq:cx-full}
\end{eqnarray}
where we have assumed that ions and neutrals have the same mass. Instead of computing this computationally expensive integral over velocity space, we have implemented a simplified model based on Refs.~\citenum{MeierThesis,Pauls1995}. The resulting collisional charge exchange terms are 
\begin{equation}
    C^{cx}_i = -C^{cx}_n = \sigma_{cx} V_{cx} (n_i f_n - n_n f_i). \label{eq:cx-in}
\end{equation}
The effective relative velocity $V_{cx}$ is defined by 
\begin{equation}
    V_{cx}^2 \equiv \frac{4}{\pi}(v_{t,i}^2 + v_{t,n}^2) + (\bm{u}_i - \bm{u}_n)^2,
\end{equation}
where $\bm{u}_s$ and $v_{t,s}^2$ are the fluid velocity and squared thermal speed for species $s$. Cross section data is taken from Ref.~\citenum{barnett}, and we currently only consider hydrogen (H-H$^+$) and deuterium (D-D$^+$) reactions using
\begin{eqnarray}
    \sigma_{cx, H} = 1.12 \times 10^{-18} - 7.15 \times 10^{-20} \ln(V_{cx}), \\
    \sigma_{cx, D} = 1.09 \times 10^{-18} - 7.15 \times 10^{-20} \ln(V_{cx}).
\end{eqnarray}
This model currently neglects the perpendicular momentum transfer from neutrals to the $E \times B$ flow. When a neutral becomes an ion via charge exchange, the location of the neutral and the guiding center of the resulting ion differ by a distance of $\rho = \v{u} \times \hat{\bhat}/\Omega_{ci}$. This process is the same as the pickup ion effect that occurs in the solar wind.\cite{isenberg1987evolution} This important feature will be implemented in future work.

As noted earlier, the plasma species and the neutral species distribution functions are evolved on the same physical space grid but different velocity space grids. Since $f_n$ appears in the ion collisional terms $C^{iz}_i$ and $C^{cx}_i$ and $f_n$ appears in $C^{cx}_n$, an interpolation scheme in velocity space is necessary to project the distribution functions onto another phase-space grid. In a discontinuous Galerkin scheme, interpolation between velocity space grids whose cell and domain boundaries do not overlap is not trivial, especially when one wishes to avoid conservation and aliasing errors as done previously.\cite{HakimJuno:2020} While complexity of this development effort is outside the scope of this work, it will be implemented in the future in order to retain all kinetic effects.

For now we avoid such a scheme by assuming that distribution functions are approximately Maxwellian. Thus, fluid moments $n_i$, $\v{u}_i = u_{\parallel,i} \hat{\bhat}$, and $v^2_{t,i}$ are taken from the ion distribution function $f_i$ in order to project a Maxwellian function $f_{M,i}(n_i,\v{u}_i,v^2_{t,i})$ onto the neutral Vlasov grid. (Here $\hat{\bhat} = \bm{B}/|\bm{B}|$ is the magnetic field unit vector.) Similarly fluid moments $n_n$, $u_{\parallel,n} = \bm{u}_n \cdot \hat{\bhat}$, and $v^2_{t,n}$ are used to project a Maxwellian function for neutrals $f_{M,n}(n_n,u_{\parallel,n},v^2_{t,n})$ onto the ion gyrokinetic grid. Then the collisional neutral interaction terms become
\begin{align}
    C^{iz}_i &= n_e f_{M,n} \langle \sigma v_e \rangle \label{eq:iz-ion-max}\\
    C^{cx}_i &= \sigma_{cx} V_{cx} (n_i f_{M,n} - n_n f_i) \label{eq:cx-ion}\\
    C^{cx}_n &= -\sigma_{cx} V_{cx} (n_i f_n - n_n f_{M,i}). \label{eq:cx-neut}
\end{align}
The models for ionization and charge exchange conserve particles, momentum and energy in the system, a property that was confirmed in one-dimensional tests with periodic boundary conditions.

\section{Neutral model verification and benchmarks}\label{sec:1d-sol}

The neutral model, as implemented in \gkyl, has been verified with analytic theory and benchmarked against the DEGAS2 Monte Carlo neutral code, which is detailed in this section. In these simulations, the gyrokinetic plasma species are static (not evolved) and only the neutrals are evolved. Simulation end times were chosen such that the neutral species achieved a steady state, verified by the fact that neutral profiles were no longer evolving. 

\subsection{Verification with analytic solution}\label{sec:verif}
To verify the neutral interaction models, we derive theoretical predictions of steady-state profiles for the neutral distribution function in one configuration space and one velocity space (1X1V) with a static background plasma. In 1X1V the steady-state neutral Vlasov equation with either charge exchange or ionization can be rewritten as 
\begin{equation}
    v_z \pd{f}{z} = -\nu \left(f(z,v_z)-g(z,v_z)\right), \label{eq:ss-theory}
\end{equation}
where $f$ is the neutral distribution function. When the RHS of Eq.~\ref{eq:ss-theory} represents charge exchange, $\nu = n_i\sigma_{cx}V_{cx}$ and $g$ is proportional to the ion distribution function 
\begin{equation}
    g(z,v_z) \equiv \frac{f_i(z,v_z)}{n_i}  \times \int d v_z' f(z,v_z').
\end{equation}
In the case of ionization, $\nu = n_e \langle \sigma v_e \rangle$ and $g = 0$. If we consider a finite symmetric domain with $z=[-L/2,L/2]$,  the general solution to Eq.~\ref{eq:ss-theory} is given by the piecewise exponential function
\begin{equation}
    f(z,v_z)= \left\{
        \begin{array}{ll}
        C_+(z,v_z)e^{-\frac{\nu}{v_z}(z+L/2)} & v_z>0 \\
        C_-(z,v_z)e^{-\frac{\nu}{v_z}(z-L/2)} & v_z<0, \label{eq:sln-theory}
    \end{array} \right .
\end{equation}
where $C_{\pm}$ can be determined from the imposed boundary conditions. A detailed derivation is given in Appendix \ref{apx:theory}. The zeroth moment of Eq.~\ref{eq:sln-theory} gives a theoretical prediction for density, which can be directly compared to numerical results.

\begin{figure}
    \centering
    \includegraphics[width=.8\linewidth]{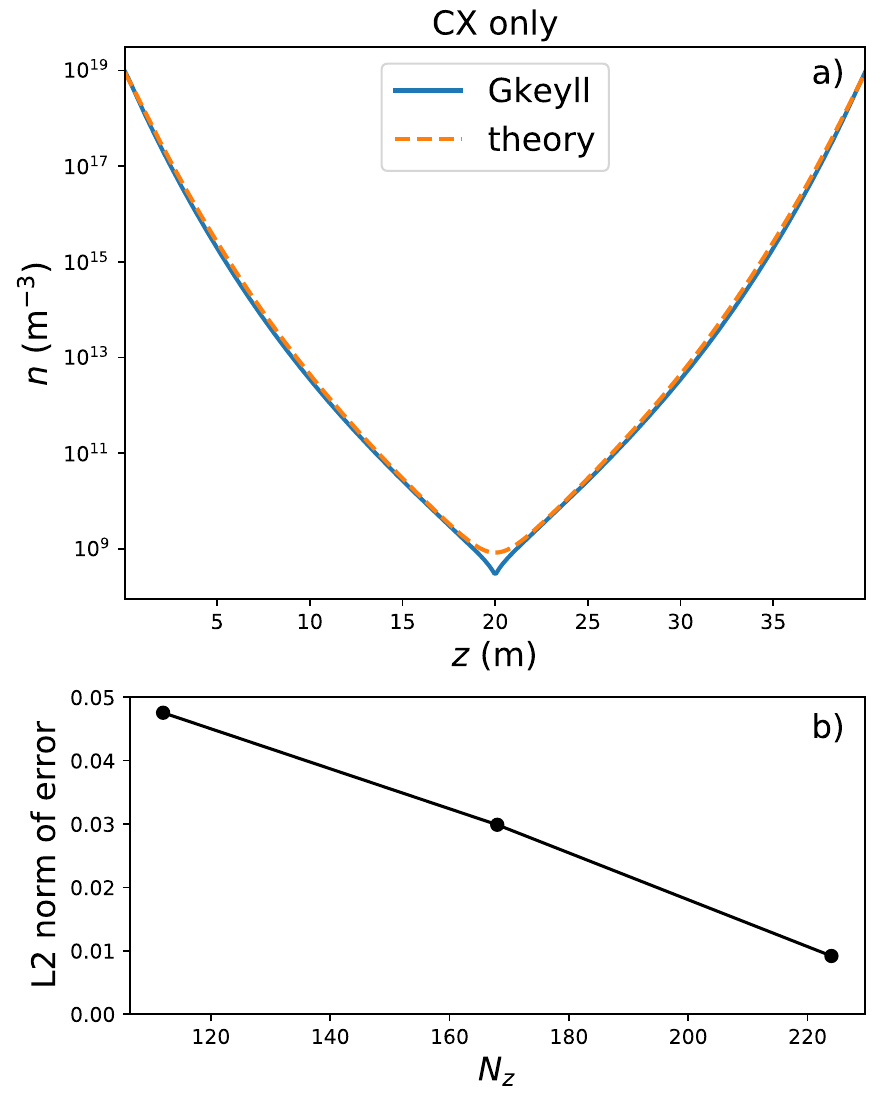}
    \caption{Verification of the charge exchange model in \gkyl. Assuming a simplified model for the charge exchange cross section (Eq.~\ref{eq:cx-neut}), the theoretically predicted steady-state density profile is compared with the resulting profile from \gkyl\ in (a). The $L^2$-norm of the error relative to the $L^2$-norm of the theoretical prediction is shown in (b).}
    \label{fig:cx-verification}
\end{figure}

To verify models of neutral interactions in \gkyl, tests were conducted in 1X1V with either charge exchange or ionization. The recycling coefficient was $\alpha_{rec} = 1.0$ and the reflection coefficient was $\alpha_b = 0$. A hydrogen plasma is assumed, and input parameters are the following: $B_0 = 0.5$ T, $n_0 = 5 \times 10^{18}$ m$^{-3}$, and $T_{e0} = T_{i0} = 20$ eV, $T_{n0}=2$ eV. The connection length is $L_z = 40$ m. The resolution is $(N_z, N_{v_z}) = (224, 32)$. The simulation was run to $t_{end} = L_z/c_{s0}$, where $c_{s0} = \sqrt{T_{e0}/m_i}$. Static plasma conditions are $n_e = n_i = n_0$, $T_e = T_i = T_{e0}$, and $\upare = \upari = -c_{s0}$ for $z \leq 0$ or $c_{s0}$ for $z > 0$.
The neutral initial conditions are 
\begin{equation}
    n_n(t=0) = \left\{
        \begin{array}{ll}
            n_{0}({\rm sech}^2(-(L_z/2 + z - 2)/0.2) + 10^{-6}) & \quad z \leq 0 \\
            n_{0}({\rm sech}^2((L_z/2 + z - 2)/0.2) + 10^{-6}) & \quad z> 0,
        \end{array}
    \right.
\end{equation}
with $T_n(t=0) = T_{n0}$ and $u_n(t=0) = 0$. Density initial conditions were chosen to approximate steady-state conditions, which include an exponential decay away from the $z$-boundaries, so that the actual steady state could be achieved more quickly in the code. We considered the case with only charge exchange and a constant value for the reaction rate $\sigma_{cx} V_{cx} = 2.2
\times 10^{-14}$ m$^3$/s. A narrow, shifted Maxwellian sourced the recycled neutrals at the boundary
\begin{equation}
    f_{M,rec} = \frac{1}{\sqrt{2 \pi v_{t,n0}^2}} \; e^{-(v_z \mp c_s)^2/(2 v_{t,n}^2)},
\end{equation}
where $v_{t,n0}^2 = T_{n0}/m_i$, and the minus and plus signs correspond to the minimum and maximum boundaries in $z$, respectively. A scaling factor $C_{rec} = 4.98 \times 10^{18}$ m$^{-3}$ was used such that the incoming neutral flux matched the incident ion flux of the static plasma. The symmetric solution for the density, Eq.~\ref{eq:sym-density}, was solved numerically and compared to the steady-state solution from \gkyl. This comparison is shown in Fig.~\ref{fig:cx-verification}a. The $L^2$-norm of the error, normalized to the $L^2$-norm of the solution predicted by theory, is shown in Fig.~\ref{fig:cx-verification}b for Gkeyll resolutions $N_z = 112, 168, 224$. The error decreases with increasing resolution.

\begin{figure}
    \includegraphics[width=.8\linewidth]{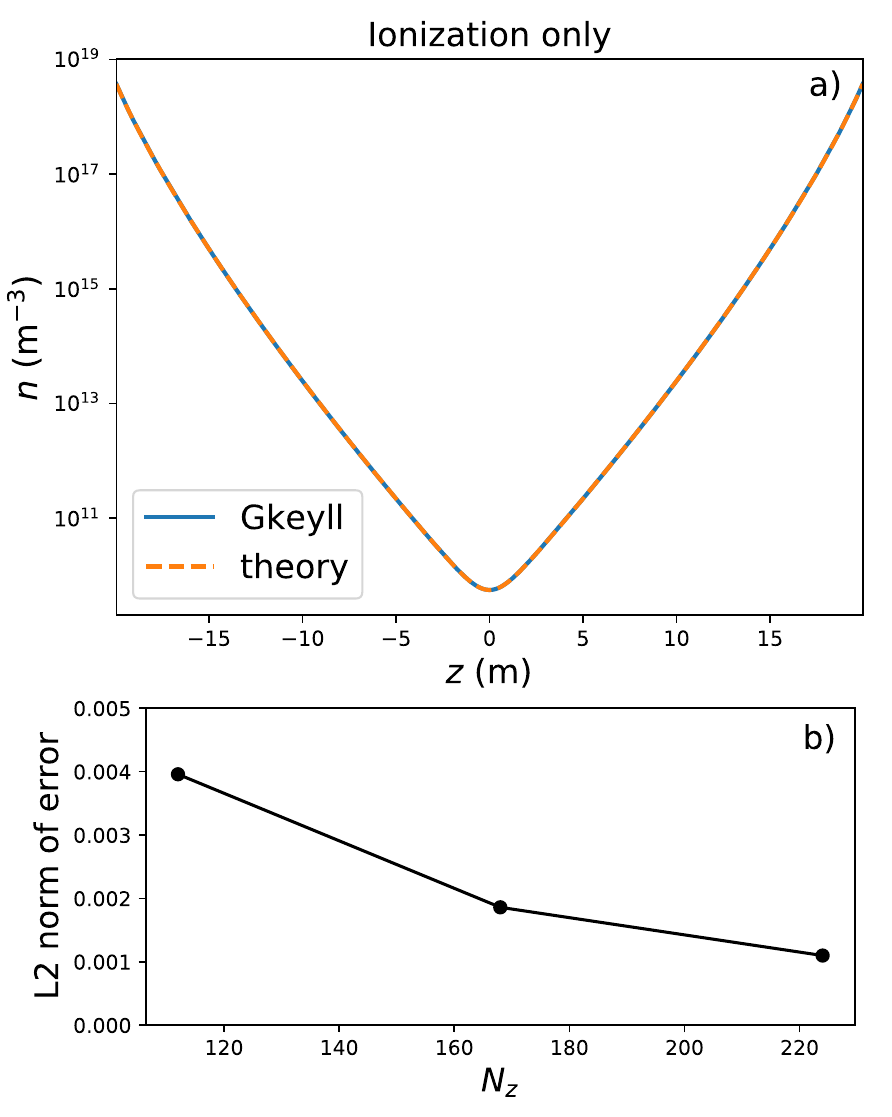}
    \caption{Verification of the ionization model in \gkyl. Assuming a simplified ionization model (Eq.~\ref{eq:iz-neut}), the theoretically predicted steady-state density profile is compared with the resulting profile from \gkyl\ in (a). The $L^2$-norm of the error relative to the $L^2$-norm of the theoretical prediction is shown in (b).}
    \label{fig:iz-verification}
\end{figure}

The same test case was run with ionization only and the resulting density was compared to the analytic solution, Eq.~\ref{eq:sym-density} with $\beta_i =0$. This comparison is shown in Fig.~\ref{fig:iz-verification}a with the associated $L^2$-norm of the error in \ref{fig:iz-verification}b, demonstrating convergence. Both cases show excellent agreement with theoretical predictions of density profiles. 

\subsection{Benchmarks with DEGAS2}\label{sec:degas}
Additional tests with static plasma species were carried out and compared with results from the DEGAS2 Monte Carlo neutral code. These tests were run with low and high densities with both ionization and charge exchange and also with each neutral interaction separately. Hydrogen ions are assumed, and input parameters are $B_0 = 0.5$ T, $n_0 = 5 \times 10^{18} \, (10^{19})$ m$^{-3}$, $T_{e0} = 30$ eV $T_{i0} = 60$ eV, $T_{n0} = 10$ eV, and $L_z = 40 \, (10)$ m, with changes to the high density parameters denoted by parentheses. The resolution is $(N_z,N_{v_x},N_{v_y},N_{v_z})=(224,16,16,16)$ for the low density case and $(N_z,N_{v_x},N_{v_y}, N_{v_z})=(448,16,16,16)$ for the high density case. Both were run to an end time of $t_{end} = 3L_z/c_{s0}$, where $c_s = \sqrt{T_{e0}/m_i}$. 

Static plasma parameters are $n_e = n_i = n_0$, $T_e = T_{e0}$, $T_i = T_{i0}$, and $\upare = \upari = \frac{c_s z}{L_x/2}$. 
The neutral initial conditions are 
\begin{equation}
    n_n(t=0) = \left\{
        \begin{array}{ll}
            n_{0}({\rm sech}^2(-(L_z/2 + z - 2)/0.2) + 10^{-6}) & \quad z \leq 0 \\
            n_{0}({\rm sech}^2((L_z/2 + z - 2)/0.2) + 10^{-6}) & \quad z > 0,
        \end{array}
    \right. \\
\end{equation}
with $T_n(t=0) = T_{n0}$ and $\bm{u}_n(t=0) = (0,0,0)$. A Maxwellian neutral source function $S_n$ with density source rate $8 \times 10^{21}$ (m$^3$/s) $n_0^2$ and temperature $T_{n0}$ is used to approximate a source floor due to recombination. Recycling wall boundary conditions were used for neutrals with a recycling coefficient of $\alpha_{rec} = 1.0$ and no reflection of neutrals ($\alpha_b = 0$). Both high and low density tests were run on 112 cores and took 44 and 183 minutes, respectively.
\begin{figure}
    \centering
    \includegraphics[width=.8\linewidth]{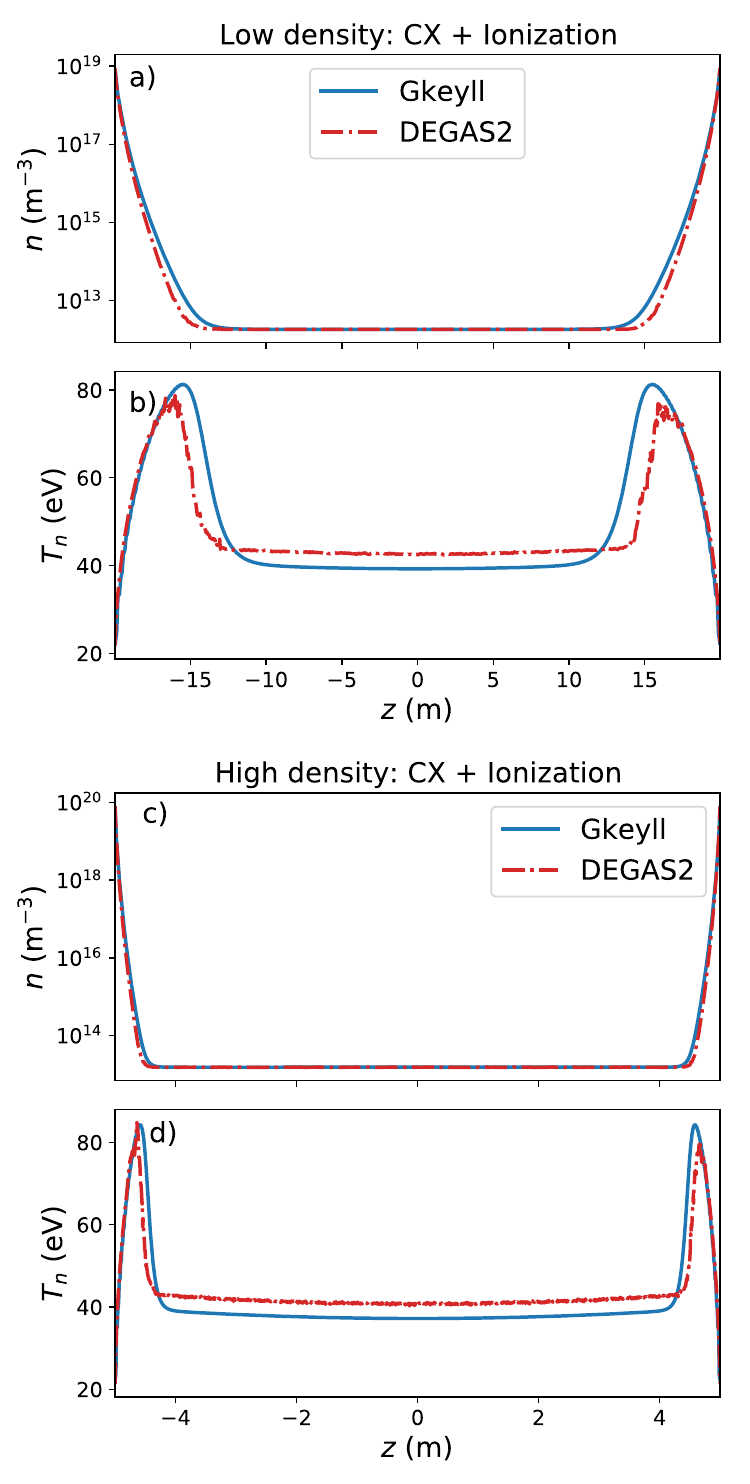}
    \caption{A comparison of benchmark tests from \gkyl\ and DEGAS with static plasma species that include charge exchange, ionization and a recycling source at the walls. Steady-state profiles of neutral density (a) and temperature (b) are shown from the low plasma density case. Neutral density (c) and temperature (d) are also compared from the high plasma density case.}
    \label{fig:gkyl-d2-both}
\end{figure}

DEGAS2 was run at identical conditions for the high density and low density benchmark cases. The main differences are the lack of higher-order information in the cells and the use of internal tables for the charge exchange cross section and ionization rates. Also, DEGAS2 is run in 3X3V instead of 1X3V, albeit with symmetry in physical space. The domain is divided into 896 and 1792 zones, wherein properties are constant, in the low and high density cases, respectively. This is higher than the \gkyl\ resolution because only cell-averages are available in DEGAS2, whereas the \gkyl\ results contain higher-order spatial information. To minimize noise in the region of low neutral density, 16M trajectories were used for each of two source types, volumetric and surface. With 128 cores, each DEGAS2 calculation took about 10-20 minutes. These tests were solving the steady-state equations, while the \gkyl\ tests dynamically evolved to a steady state. This comparison focuses on benchmarking the model in \gkyl\ and not on performance; thus, no efforts were made to optimize either DEGAS2 or \gkyl\ for these tests. 
\begin{figure}
    \centering
    \includegraphics[width=.8\linewidth]{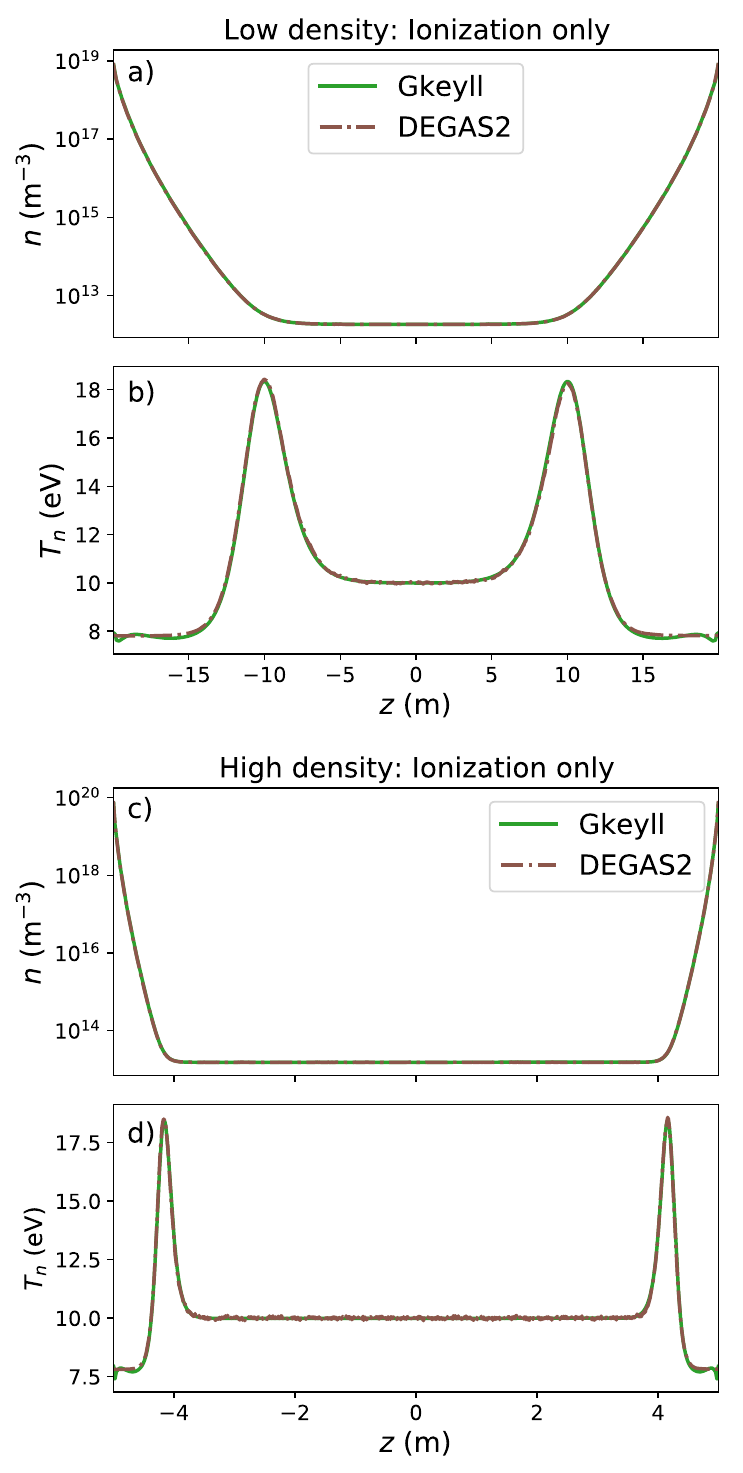}
    \caption{A comparison of benchmark tests from \gkyl\ and DEGAS with static plasma species that include only ionization and a recycling source at the walls. Steady-state profiles of neutral density (a) and temperature (b) are shown from the low plasma density case. Neutral density (c) and temperature (d) are also compared from the high plasma density case.}
    \label{fig:gkyl-d2-iz}
\end{figure}

Since the ionization reaction rate is constant in time and space, the values calculated from the DEGAS2 model were used in the \gkyl\ cases. In DEGAS2, only the ground state is transported, and the effect of excitation is taken into account in the effective rate that was used for both codes. This was $\langle v_e \sigma \rangle_{iz} = 2.147 \times 10^{-14}$ m$^3$/s. Figure \ref{fig:gkyl-d2-both} compares the \gkyl\ and DEGAS2 results for the cases including both ionization and charge exchange. Steady-state neutral density profiles are shown in figures \ref{fig:gkyl-d2-both}a and \ref{fig:gkyl-d2-both}c, demonstrating a slightly larger penetration length for neutrals in \gkyl. Neutral temperature profiles are compared in figures \ref{fig:gkyl-d2-both}b and \ref{fig:gkyl-d2-both}d. Relative to DEGAS2, \gkyl profiles differ by up to 40\% near the temperature peaks and less than 10\% elsewhere. Overall agreement is better for the higher density case, seen in Figs.~\ref{fig:gkyl-d2-both}c and \ref{fig:gkyl-d2-both}d. In both DEGAS2 and \gkyl, the neutral temperature peaks around 80 eV at the radial location where the density approaches a minimum value. This demonstrates that at high densities neutrals cannot penetrate far before becoming ionized, except for the more energetic particles. Since both codes are kinetic, the neutrals can attain maximum temperature values above the mean temperatures of the static ions and electrons with which they interact (60 eV and 30 eV, respectively). In the center of the $z$-domain, the neutrals are born from the source floor with a temperature of 10 eV. Most are ionized before they can equilibrate to the 60 eV ion temperature via the charge exchange process; hence the temperature is close to 40 eV.

For charge exchange, DEGAS2 uses the integral collision operator (Eq.~\ref{eq:cx-full}) with total cross sections from Refs.~\citenum{Janev1993atomic,krstic1998atomic}, while \gkyl\ uses a fitting function to approximate the cross section that depends only on moments of the distribution function. When only ionization is included and both codes use the same ionization rate, near perfect agreement results as seen in Fig.~\ref{fig:gkyl-d2-iz}, indicating that differences in Fig.~\ref{fig:gkyl-d2-both} arise from the charge exchange model. Without the charge exchange process, whereby neutrals gain energy from the ions, the neutral temperature is much lower, the peaks due to the penetration of high energy neutrals is again visible at the point where the density profiles approach the minimum value. 

\begin{figure}
    \centering
    \includegraphics[width=.8\linewidth]{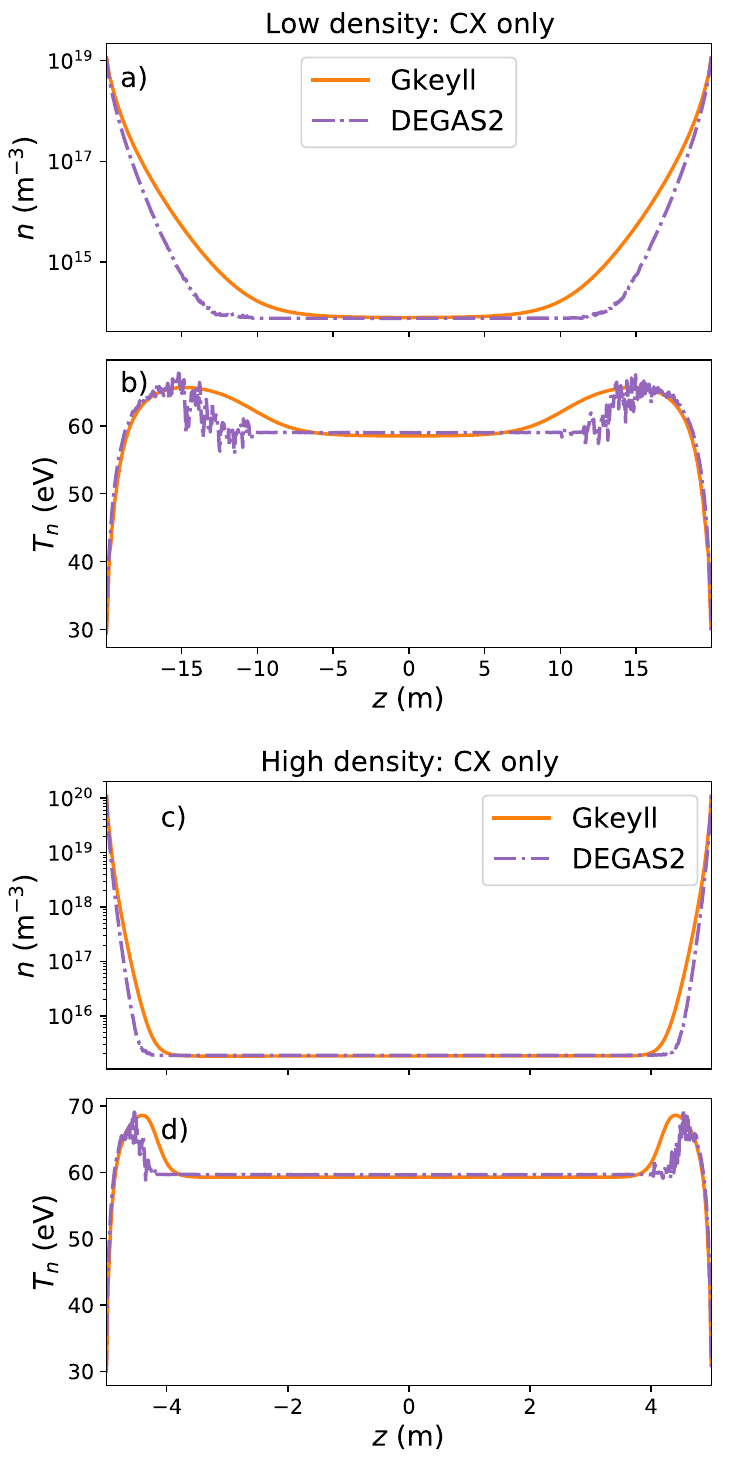}
    \caption{A comparison of benchmark tests from \gkyl\ and DEGAS with static plasma species that include only charge exchange and a recycling source at the walls. Steady-state profiles of neutral density (a) and temperature (b) are shown from the low plasma density case. Neutral density (c) and temperature (d) are also compared from the high plasma density case.}
    \label{fig:gkyl-d2-cx}
\end{figure}

The comparison for tests including only charge exchange is shown in Fig.~\ref{fig:gkyl-d2-cx}. In \ref{fig:gkyl-d2-cx}a the \gkyl\ penetration length is noticeably greater than that in DEGAS2 and the temperature peak is broader. Relative to DEGAS2, \gkyl profiles differ by approximately 10\% near the temperature peaks. Agreement is better in the high-density comparison, shown in Fig.~\ref{fig:gkyl-d2-cx}c-d, since the DEGAS2 charge exchange operator converges towards a statistical average in the limit of high collisionality. The \gkyl\ model of charge exchange is most appropriate when neutral densities are greater than $10^{17}$ m$^{-3}$, since this is where profiles approach DEGAS2 values. This comparison highlights the importance of an integral collision operator for the charge exchange model. Including this in \gkyl\ is computationally feasible, particularly if a lower order form of the distribution function is assumed, i.e. Maxwellian. This will be addressed in future development work. 

\section{Scrape-off layer simulations using the coupled plasma-neutral model}\label{sec:nstx}
The neutral model, which evolves neutrals in three physical space plus three velocity space dimensions (3X3V) has been tested in a coupling with gyrokinetic simulations of the NSTX SOL, in three physical space dimensions and two velocity space dimensions (3X2V). These have been extended from previous \gkyl\ NSTX simulations\cite{shi2019full,Shi2017thesis,mandell2020electromagnetic,Mandell2021thesis} and compared to a baseline case without neutrals. In this section we describe the setup and results of the simulations.

\subsection{Simulation setup} \label{sec:setup}
In the baseline case, which is labeled as ``no neutrals'' in the plots that follow, electron and ion species were evolved with the electrostatic gyrokinetic model, Eqs.~\ref{eq:gk}-\ref{eq:poisson}, in three configuration space dimensions and two velocity space dimensions (3X2V). We assume a simplified helical magnetic geometry, with only open field lines and constant field line length and curvature throughout the domain, which is described in more detail in Appendix \ref{apx:geo}. We approximate NSTX SOL conditions for an L-mode discharge, assuming a singly-ionized deuterium plasma and power into the SOL $P_{SOL} = 1.35$ MW. Relevant parameters for the magnetic geometry include the magnetic field on axis $B_{axis} = 0.5$ T, the major radius at the magnetic axis $R_0 = 0.85$ m, and the minor axis $a_0 = 0.5 $ m. Plasma temperatures $T_{e0} = T_{i0} = 40$ eV are used to set the velocity grid extents and initial conditions.

\begin{figure*}
    \centering
    \includegraphics{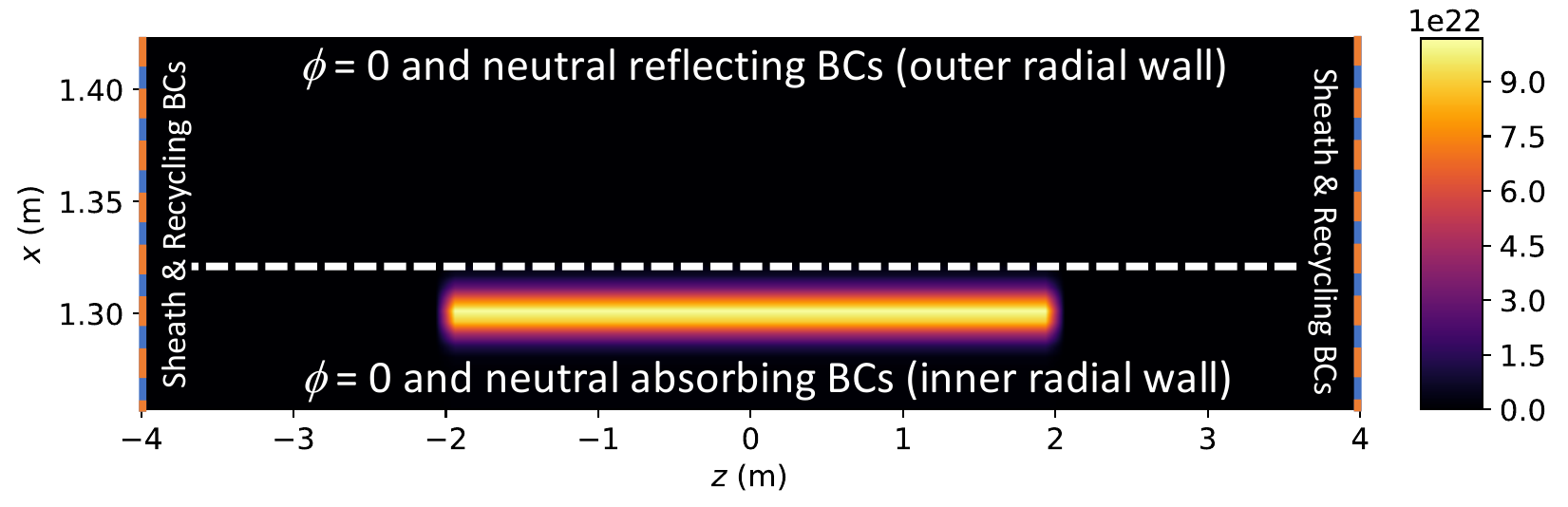}
    \caption{Plot of the density particle source (m$^{-3}$/s) for NSTX SOL simulations, shown in the $(x,z)$ plane. The dashed white line indicates the ``quasi-separatrix'' in our open-field line simulations. The dashed orange and blue line indicates the conducting sheath and recycling boundary conditions (BCs) applied in $z$ for plasma and neutral species, respectively. In the radial direction,  Dirichlet BCs with $\phi = 0$ are applied for plasma species. For neutral species, reflecting BCs are applied at the outer radial boundary ($x_{\max}$) and absorbing BCs are applied at the inner radial boundary ($x_{\min})$.}
    \label{fig:nstx-src-bc}
\end{figure*}

Perpendicular dimensions of the simulation box are $L_x = 50 \rho_{s0} \approx 14.6$ cm and $L_y = 100 \rho_{s0} \approx 29.1$ cm, where $\rho_{s0} = c_{s0}/\Omega_{ci}$, and the box is centered at $R_c = R_0 + a_0$. The parallel length of the box is $L_z = L_{pol}/\sin \chi = 8$ m, where $L_{pol} = 2.4$ m and $\chi = \sin^{-1} (B_v/B)$ is the pitch angle determined by the ratio of the vertical field magnitude $B_v$ to the total field magnitude $B$. Since the flux tube simulation box covers a fraction of the SOL, the source  is scaled as $P_{src} = P_{SOL} L_y L_z / (2 \pi R L_{pol}) = 0.31$ MW. A uniform grid is used for all species. Resolution for the gyrokinetic species is ($N_x$, $N_y$, $N_z$, $N_{\vpar}$, $N_\mu$) = (16, 32, 32, 12, 6). Convergence was checked in $k_x$ space, and the resolution is appropriate. The configuration space extents are $x \in [R_c - L_x/2, R_c + L_x/2]$, $y \in [-L_y/2,L_y/2]$, and $z \in [-L_z/2,L_z/2]$. Velocity space extents are $\vpar \in [-4v_{t,s0},4v_{t,s0}]$ and $\mu \in [0,6T_{s0}/B_0]$, where $v_{t,s0} = \sqrt{T_{s0}/m_s}$ and $B_0 = B_{axis} R_0/R_c$. In the discontinuous Galerkin scheme, solutions are represented using piecewise-linear $(p=1)$ basis functions on each cell, which effectively doubles the resolution of the data in each direction. Dirichlet boundary conditions with $\phi = 0$ are applied in the radial direction and periodic boundary conditions are used in the binormal direction. Conducting-sheath boundary conditions\cite{shi2015gyrokinetic,shi2017gyrokinetic} are applied in the parallel direction, which reflect low-energy electrons back into the domain. Ions free stream out of the domain along $z$. The location of these boundary conditions is shown in Fig.~\ref{fig:nstx-src-bc}.

Plasma species are sourced with non-drifting Maxwellian distribution functions to model a particle and heat source at the midplane. The particle density source is defined as 
\begin{equation}
    S_n(x,z)= \left\{
        \begin{array}{ll}
        S_0 \max\left[\exp \left( \frac{-(x-x_S)^2}{2 \lambda_S}\right), 0.01 \right], & |z|<L_z/4 \\
        0 & {\rm else}, \label{eq:src-dens}
    \end{array} \right.
\end{equation}
where $x_S = R_c - 0.05$ m and $\lambda_S = 0.005$ m and is shown in Fig.~\ref{fig:nstx-src-bc} in the $(x,z)$ plane. The particle source rate $S_0$ is calculated for a given source temperature to achieve the desired power $P_{src}/2$, since power is divided equally between ions and electrons. The temperature source was increased relative to previous NSTX \gkyl\ simulations to account for heat loss due to ionization and charge exchange interactions. Due to discretization the actual source temperature ended up being $T_{e,i} \approx 150$ eV for $x < x_S+ 4\lambda_S$ and $T_{e,i} \approx 130$ eV for $x \geq x_S+ 4\lambda_S$. The plasma species were initialized to a Maxwellian, with density defined similarly to the source density function, Eq.~\ref{eq:src-dens} and a temperature of $T_{e,i} = 50$ eV for $x < x_S + 3 \lambda_S$ and $T_{e,i} = 20$~eV for $x \geq x_S + 3 \lambda_S$. In Fig.~\ref{fig:nstx-src-bc} the ``quasi-separatrix'' is denoted by the white dashed line, which separates the source region from the SOL region. Note that this is fixed at a radial value of $x\approx 1.32$ m along the field line. This is lower than the typical location of the separatrix at the outer midplane in the experiment ($x \approx$ 1.45 to 1.5 m).\cite{stotler2015midplane,scotti2021outer}

The Lenard-Bernstein, or Dougherty, collision operator was used to model plasma collisions\cite{Francisquez2020lbo} that include same and cross-species collisions. A Spitzer collision frequency is calculated from the user-defined background density $n_0 = 7 \times 10^{18}$~m$^{-3}$ and temperature $T_{e0} = T_{i0} = 40$~eV and is constant in space and time. The collision frequency was reduced by a factor of 0.01 to reduce simulation run time, since the timestep is largely dependent on the plasma collisionality. Reduced pitch-angle scattering reduces the amount of high-energy electrons that enter the sheath and could thus increase the sheath potential. Reduced resistivity can also influence blob dynamics. It pushes the simulation towards the sheath-limited regime, in which currents in the blobs are closed through parallel currents to the sheath.\cite{krasheninnikov2008recent} The collision frequency can be calculated dynamically based on local density and temperatures, but this feature was not employed for these simulations. The background density $n_0$ is also used for the ion guiding-center density $n_{i0}^{g}$ that appears in the ion polarization term of the gyrokinetic Poisson equation \ref{eq:poisson}. 

The case with neutrals is identical to the baseline case described above but with the addition of neutrals species evolved with the Vlasov model, Eq.~\ref{eq:vlasov}. Ionization and charge exchange interactions are included. The neutral grid resolution is ($N_x$, $N_y$, $N_z$, $N_{v_x}$, $N_{v_y}$, $N_{v_z}$) = (16, 32, 32, 6, 6, 6), with the same configuration space extents as the gyrokinetic species and velocity space extents $v_{x,y,z} \in [-4v_{t,i0},4v_{t,i0}]$ defined in terms of the ion thermal speed. Piecewise-linear basis functions are also used for the numerical solution of the neutral distribution function. 

Boundary conditions for neutrals are displayed in Fig.~\ref{fig:nstx-src-bc}. In the radial direction, absorbing boundary conditions are applied at the inner boundary to model neutrals that penetrate into the core, and perfectly reflecting boundary conditions are applied at the outer boundary. Since there is no ion flux at the outer radial boundary in our model, the recycling boundary conditions are not applied at the outer boundary. Periodic boundary conditions are applied in the binormal direction, and recycling boundary conditions are applied in the parallel direction. A recycling rate of $\alpha_{rec} = 0.95$ and temperature $T_{rec} = 10$ eV are assumed. 

A source floor is applied via a non-drifting Maxwellian distribution function. A spatially constant particle source rate ${S_n = 8\times10^{-21} \, {\rm m}^{3} {\rm s}^{-1} \, n_{i0}^2}$ approximates recombination, since we do not yet include a model for this process in the code. Neutrals are also initialized with a Maxwellian, with density profile that decays exponentially away from the $z$-boundaries to approximate steady-state profiles and a spatially constant temperature of 10 eV.

A discontinuous Galerkin algorithm that guarantees positivity of the distribution function and maintains conservation properties has been developed for the gyrokinetic solver\cite{Mandell2021thesis} in \gkyl\ but is not yet available for the Vlasov solver. When recycling boundary conditions are used and no positivity correction is applied, steep gradients develop in the neutral distribution function in $z$ as neutrals enter the domain, which results in unphysical, negative values for the neutral density and temperature. Therefore, a non-conservative positivity correction is currently employed for neutrals, which sets the distribution function to zero in a cell where the cell average of the numerical solution has become negative. This can result in an increase of approximately 10--20\% of neutral particles and energy relative to conservative simulations without the positivity correction. However, the unphysical negative values of density and temperature present in the latter make them unreliable baselines. A conservative positivity-preserving algorithm for the Vlasov solver is under development and will be included in future work.

In summary, the nonphysical assumptions in these proof-of-concept simulations include a collision frequency based on constant values of density and temperature, which is scaled by 0.01; a temperature of 10 eV for recycled neutrals; and the non-conservative positivity correction for the neutral distribution function.  

\subsection{Results} \label{sec:nstx-results}

The simulations were run to 0.4 ms, which is about 4 ion transit times, $\tau_i = L_z/(2c_{s0})$. Density profiles were still evolving since the sources and sinks were not yet balanced. Changes to the profiles are less than $ \lesssim 10\%$ on relevant timescales ($\sim 10$ $\mu$s) based on neutral collisions and turbulence correlation times. Coulomb collisions were not considered due to the artificially reduced collisionality, which was described in the previous section. On the Cori Haswell nodes at NERSC, the case without neutrals required approximately 18.5k CPU-hours and the case with neutrals required approximately 93.4k CPU-hours, or about 5 times as many resources. 

\begin{figure}
    \centering
    \includegraphics[width=\linewidth]{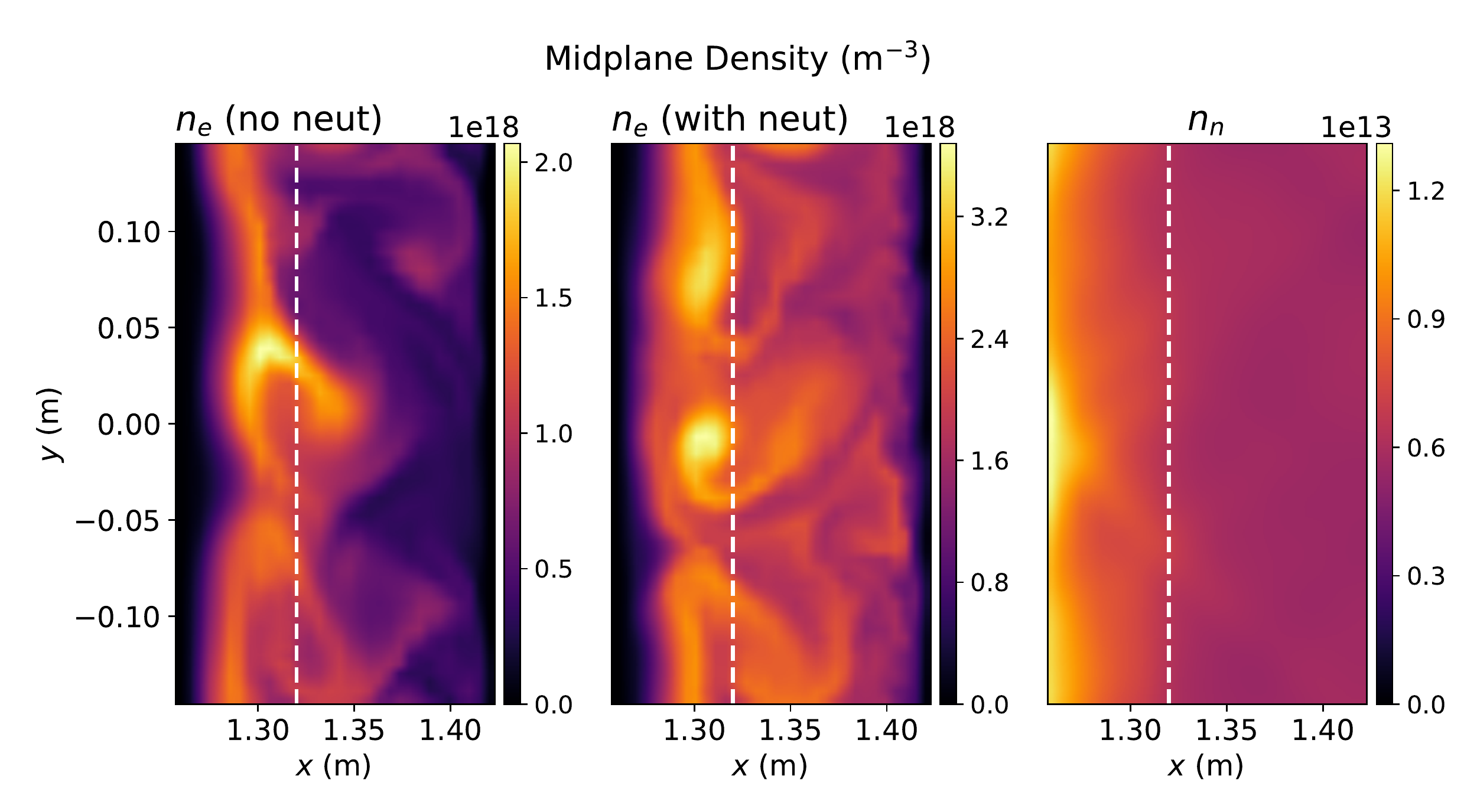}
    \caption{Snapshots of density taken perpendicular to the magnetic field at the midplane. In all plots, a dashed white line represents a ``quasi-separatrix'' that separates the source region from the SOL region. The left plot depicts electron density from the simulation without neutrals. The center plot depicts electron density from the simulation with neutrals, and the right plot shows the neutral density.}
    \label{fig:n-comp}
\end{figure}

Snapshots of midplane densities are shown in Fig.~\ref{fig:n-comp}, with the electron density from the case without neutrals shown on the left, electron density from the case with neutrals shown center, and the neutral density shown on the right. The electron density without neutrals displays a sharper density gradient, while the case with neutrals displays a density that is more constant radially. The magntidues are close to NSTX experimental values for L-mode discharges (see Fig.~7 in Ref.~\citenum{scotti2021outer}). The neutral density is five orders of magnitude less than the plasma density. Since we do not include recycling boundary conditions at the outer radial wall (see Fig.~\ref{fig:nstx-src-bc}), neutrals are only sourced at the endplates and most are ionized prior to reaching the midplane. Because of this simulation neutral values at the midplane are significantly less than in experiment.\cite{stotler2015midplane,scotti2021outer} The neutral profile is approximately constant radially in the SOL region, whereas the neutral profile decays exponentially from the outer wall to lower radii. In the simulation, neutral density is artificially high on the inner radial boundary due the conducting wall ($\phi = 0$) radial boundary conditions used for the plasma species. Because of this, the plasma density drops at the inner radial boundary, which results in lower ionization rates. When closed flux surface geometry is eventually included, plasma density will be higher inside the separatrix, and that feature should vanish. 

\begin{figure}
    \centering
    \includegraphics[width=\linewidth]{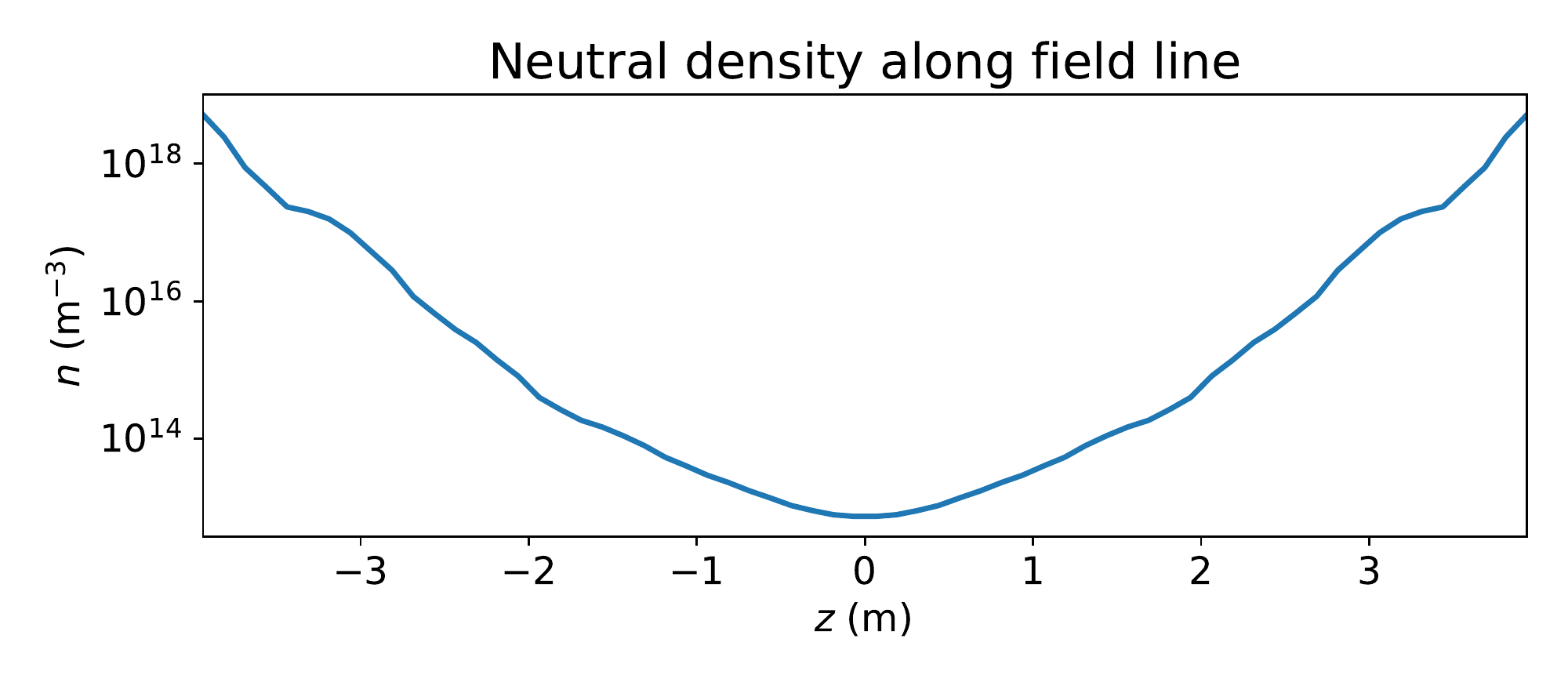}
    \caption{Neutral density profile along $z$, averaged in $x$, $y$ and time.}
    \label{fig:nn-zprof}
\end{figure}

Including recycling at the outer radial boundary will likely improve comparison with experiment. Recycling is included at the $z$-boundaries and neutral densities at these boundaries, shown by the profile in $z$ in Fig.~\ref{fig:nn-zprof}, approach experimental values.\cite{stotler2015midplane,scotti2021outer}
\begin{figure}
    \centering
    \includegraphics[width=.8\linewidth]{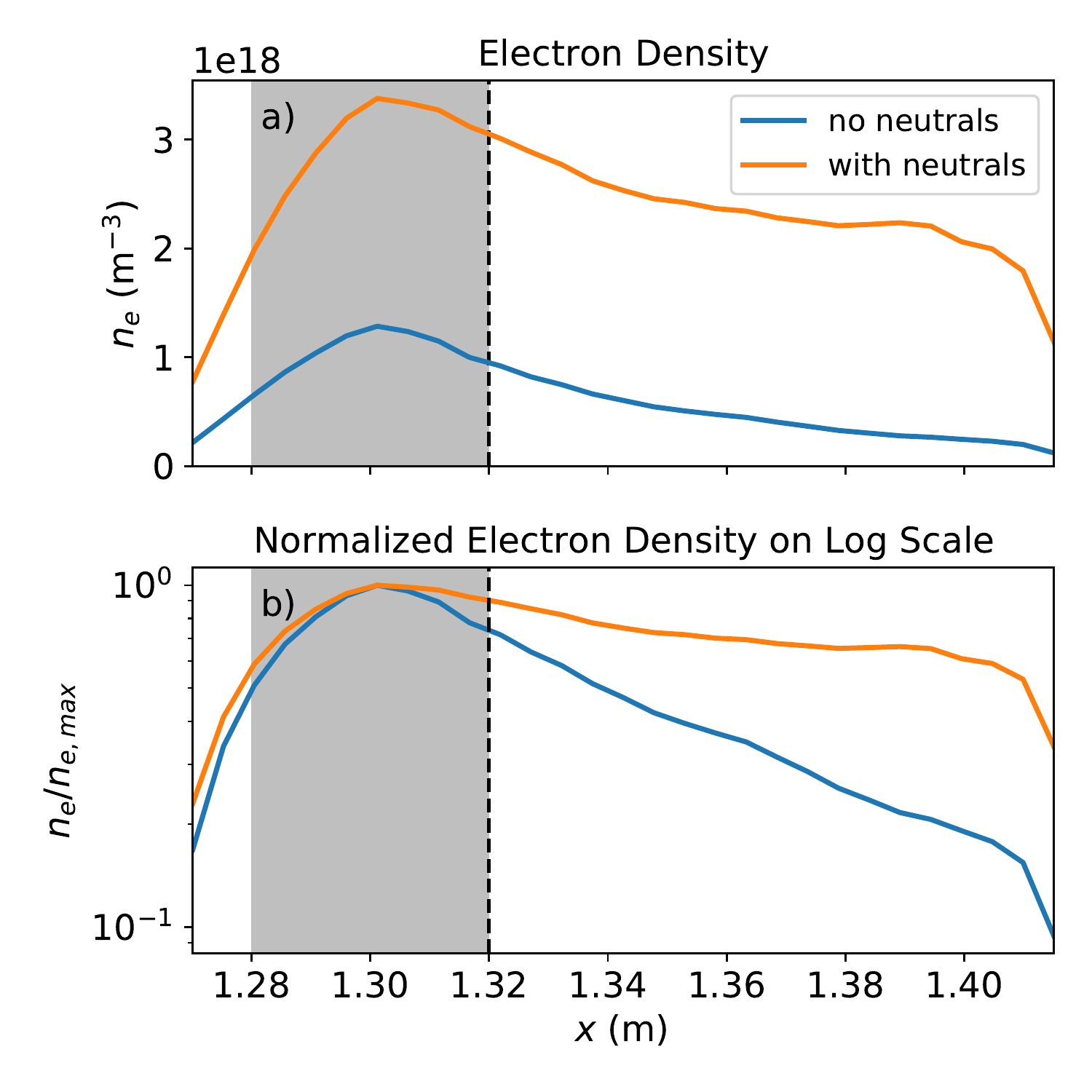}
    \caption{Steady-state density profiles are compared. In both plots the black dashed line represents the ``quasi-separatrix'' and the gray region denotes the location of the source. Plot (a) compares the electron densities from the simulations without and with neutrals on a linear scale. Plot (b) compares electron densities normalized to their maximum density on a log scale to highlight the flatter density profile in the case with neutrals.}
    \label{fig:ne-shoulder}
\end{figure}
\begin{figure}
    \centering
    \includegraphics[width=.8\linewidth]{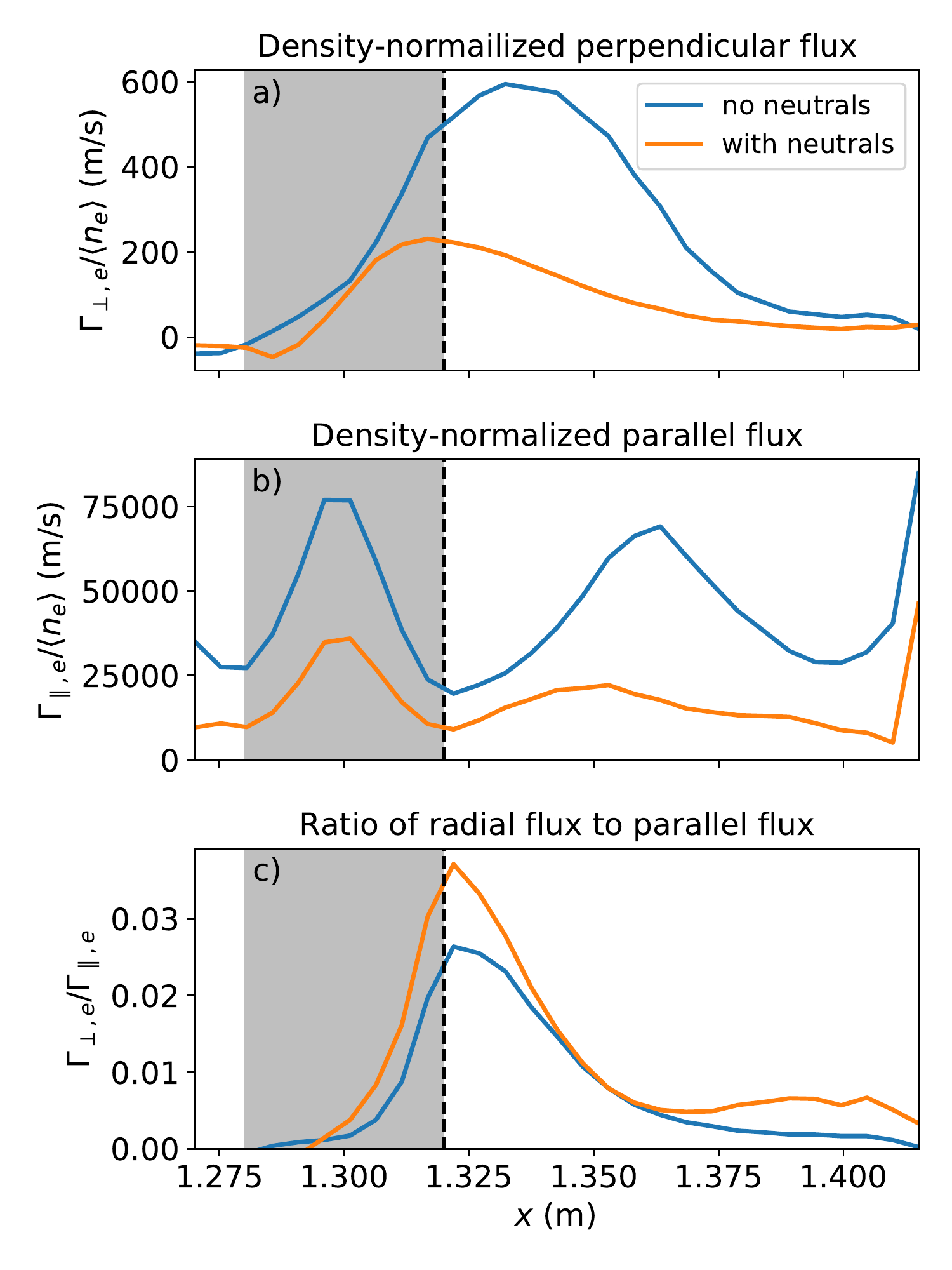}
    \caption{Radial turbulent fluxes, which are calculated as the product of density fluctuations and radial \ExB\ velocity fluctuations and normalized to the mean density, are compared in (a).  Electron parallel particle fluxes, which are calculated from the electron distribution functions at the $z$ boundary and normalized to mean density, are compared in (b). The ratio of the fluxes are compared in (c).}
    \label{fig:gx-gz}
\end{figure}
Steady-state density profiles were obtained by averaging fields from 0.3 to 0.4 ms. Electron density profiles at the midplane are compared in Fig.~\ref{fig:ne-shoulder}. The source region is shaded gray and the ``quasi-separatrix'' is denoted by the vertical dashed line. In the simulation with neutrals, ionization increases density by about a factor of 3, as seen in Fig.~\ref{fig:ne-shoulder}a. We also observe a flattening of the density profile to the right of the peak for the case with neutrals, which is seen most clearly in Fig.~\ref{fig:ne-shoulder}b, where we have normalized each density profile to its maximum value and then plotted them on a log scale. This appears similar to the density shoulder that has been observed experimentally \cite{Boedo2003,Rudakov2005,Carralero2017,Vianello2017,Vianello2019} and also studied in recent GBS simulations that include neutral interactions.\cite{Mancini2021} However, additional \gkyl\ simulations with neutrals and scans in the density source are necessary to make direct comparisons to those results.

In Figure \ref{fig:gx-gz}, we compare the perpendicular and parallel particle fluxes, normalized to the density profiles (from Fig.~\ref{fig:ne-shoulder}a). The radial perpendicular flux is due mostly to turbulence and is thus calculated from the density fluctuations and the radial \ExB\ velocity fluctuations, $\Gamma_\perp = \langle \tilde{n}_e \tilde{v}_r\rangle_{y,t} $, where fluctuating quantities are given by $\tilde{A} = A - \langle A \rangle_{y,t}$. The radial velocity is given by $v_r = -(1/B)\partial{\phi}/\partial{y}$. In Fig.~\ref{fig:gx-gz}a, the normalized perpendicular flux is reduced in the simulation with neutrals. The peak is shifted to higher radii, and the gradient of the flux profile is shallower. The normalized parallel flux is compared in Fig.~\ref{fig:gx-gz}b and is calculated as
\begin{equation}
    \Gamma_{\parallel,e} = \left\langle \left. \int \jac f_{e\,h} \dot{\bm{R}}_h \cdot \hat{\bhat} \, {\rm d}x \, {\rm d}y \, {\rm d}^3 \, \bm{v} \right|_{z=L_z/2} \right\rangle.
\end{equation}
The subscript $h$ on denotes that only electrons with kinetic energy large enough to overcome the potential barrier at the pre-sheath are included in this calculation. This is consistent with the conducting sheath boundary conditions described in the previous section. The normalized parallel flux to the endplates decreases for the case with neutrals; the parallel outflow is slower since the electron temperature has decreased. In Fig.~\ref{fig:gx-gz}c, the ratios of the radial to the parallel fluxes are compared. This ratio for the neutrals case is generally larger than the baseline simulation, suggesting that the relative decrease in parallel flow leads to density profile flattening, similar to GBS simulations with neutrals.\cite{Mancini2021}

\begin{figure}
    \centering
    \includegraphics[width=.8\linewidth]{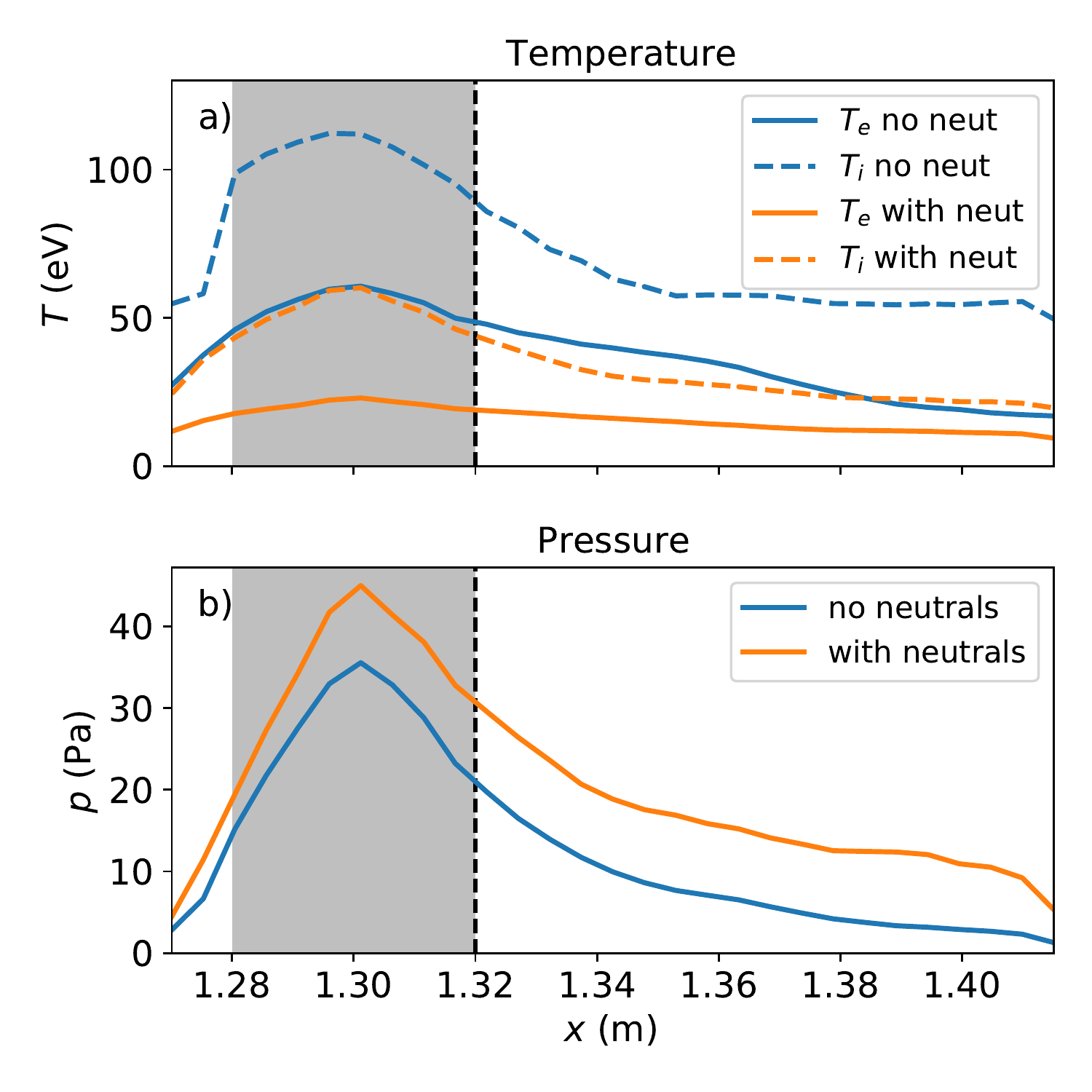}
    \caption{Steady-state temperature profiles of electron and ion temperature are compared in (a). Pressure profiles are compared in (b).}
    \label{fig:t-p}
\end{figure}
Midplane steady-state temperature profiles are shown in Fig.~\ref{fig:t-p}a. In the simulation with neutrals, electron and ion temperatures are both reduced. The ionization energy requirement reduces the electron temperature. Ion temperature decreases due to charge exchange and interaction with electrons by cross-species collisions. The electron temperature gradient is reduced in the case with neutrals, and the electron and ion temperature profiles have more similar gradients than in the case without neutrals, indicating a stronger coupling between the two. The pressure profiles, where $p=n_e(T_e+T_i)$, are shown in Fig.~\ref{fig:t-p}b. The case with neutrals is larger due to the increase in density. 

\begin{figure}
    \centering
    \includegraphics[width=.8\linewidth]{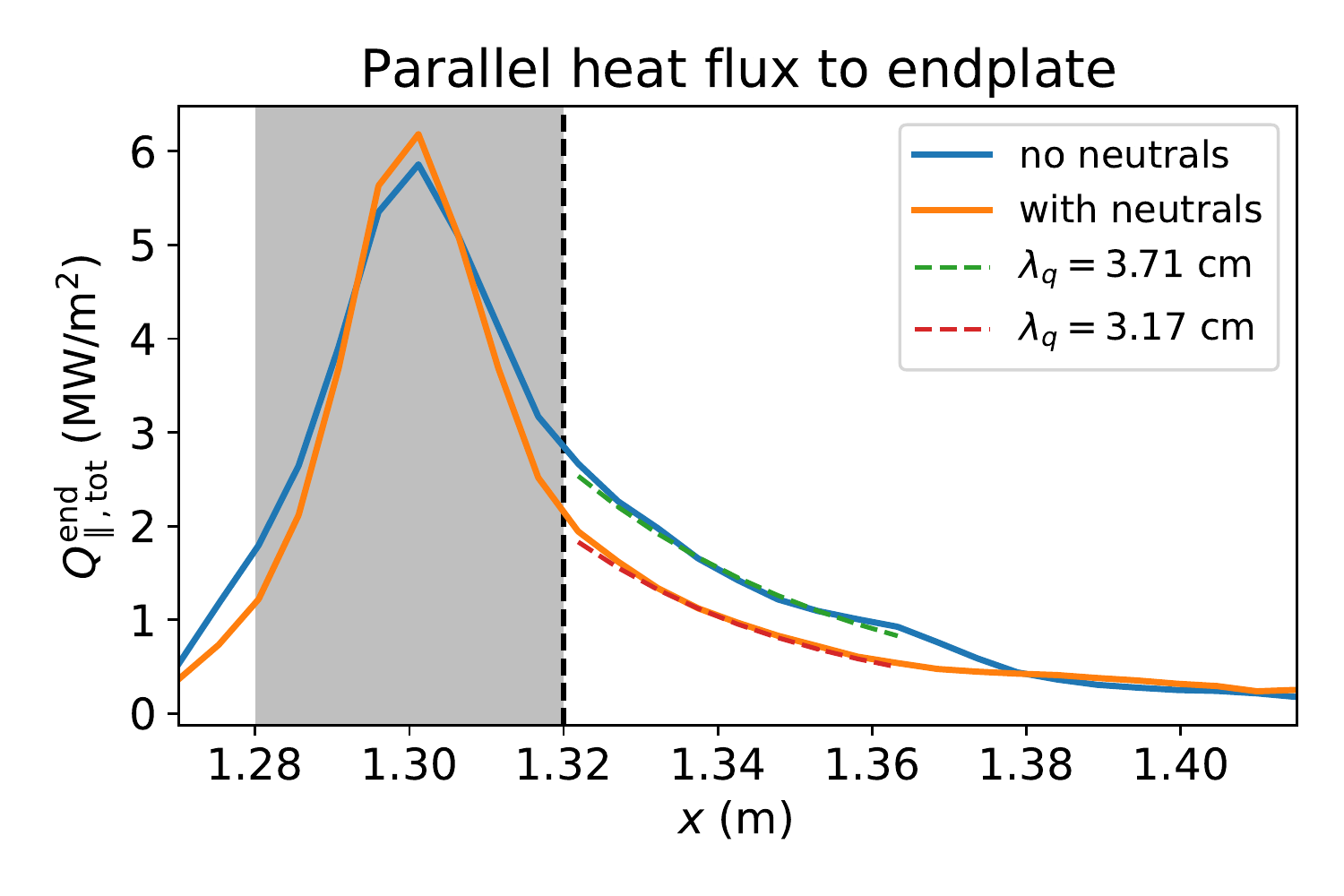}
    \caption{Parallel heat fluxes to the endplates are compared. These profiles are sums of electron and ion contributions, which were calculated from the distribution functions using Eq.~\ref{eq:qpara}. Heat flux widths were estimated from exponential fits, $\exp(-x/\lambda_q)$, at the ``quasi-separatrix'', shown by the green and red dashed curves.}
    \label{fig:qpara}
\end{figure}

The profiles of parallel heat flux to the endplates are compared in Fig.~\ref{fig:qpara}. The heat flux for each species is calculated as 
\begin{equation}
    Q_{\parallel, s}^{\rm end} = \left\langle \left. \int H_{s\, h} \jac f_{s \,h} \dot{\bm{R}}_h \cdot \hat{\bhat} \, {\rm d}x \, {\rm d}y \, {\rm d}^3 \, \bm{v} \right|_{z=L_z/2} \right\rangle, \label{eq:qpara}
\end{equation}
and the total $Q_{\parallel, \rm tot}^{\rm end} = Q_{\parallel, e}^{\rm end} + Q_{\parallel, i}^{\rm end}$ is plotted. We estimate the width of the profiles with an exponential fit just outside the quasi-separatrix, $Q_{\parallel}^{\rm end}(x=1.32)e^{-x/\lambda_q}$, and the value for the case with neutrals is slightly less than the case without neutrals. 

\begin{figure}
    \centering
    \includegraphics[width=.8\linewidth]{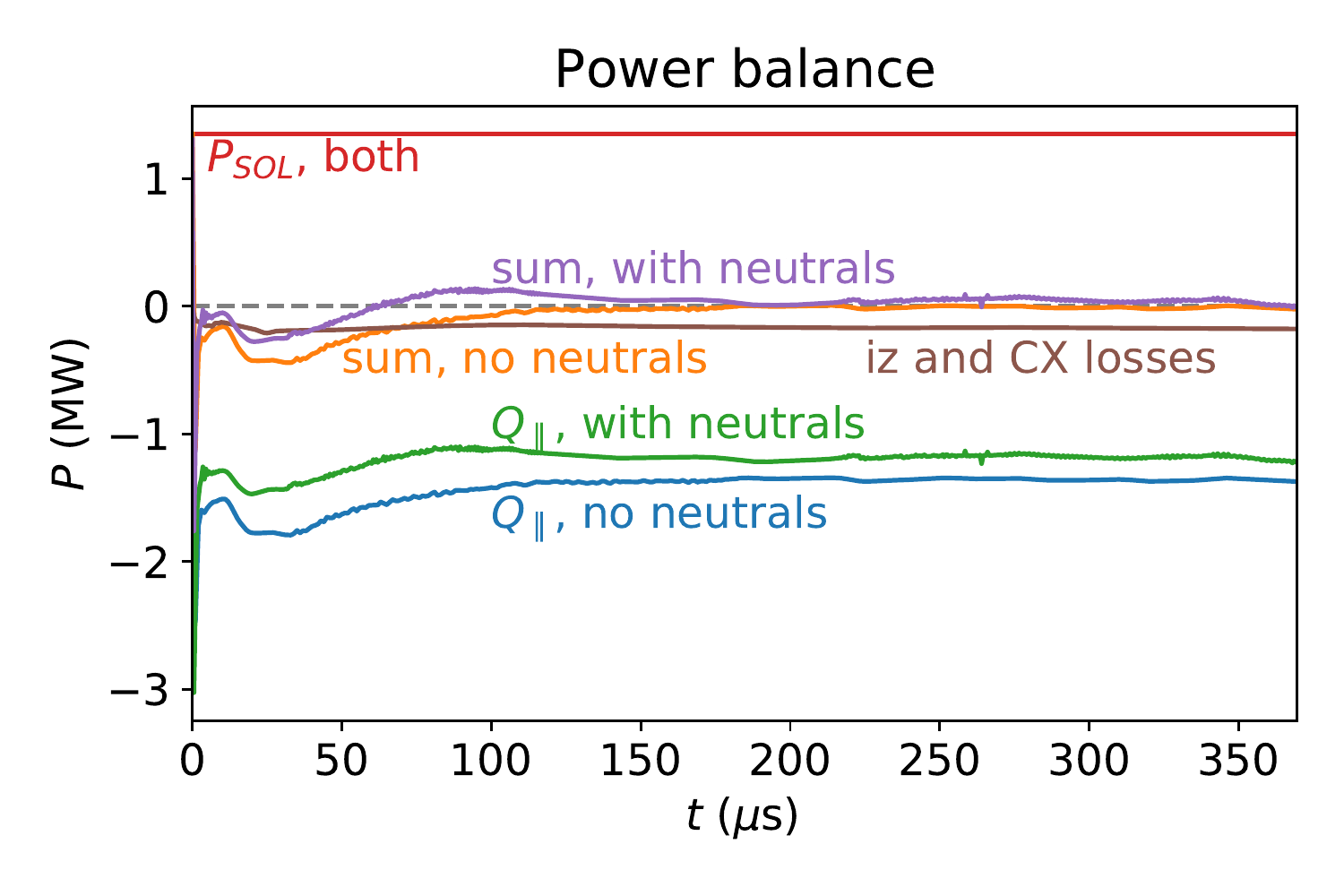}
    \caption{The power balance is calculated for the simulations without and with neutrals. Both have the same input power $P_{SOL}$ indicated by the red curve. The only power loss in the simulation without neutrals is due to heat flux out through the endplates (blue curve). The sum of these gives the orange curve. The simulation with neutrals also includes some power loss due to ionization and charge exchange (brown curve). Power loss to the endplates is given by the green curve and with the power input and other losses this sums to the purple curve.}
    \label{fig:pow-bal}
\end{figure}

The power balance in each simulation is also calculated and shown in Fig.~\ref{fig:pow-bal}. Both have the same input power at the midplane, indicated by the red line. In the case without neutrals, the only loss is at the endplates, indicated by the blue line. The sum of of the input power and loss to endplates is given by the orange line, which approaches zero. This conforms to expectations for \gkyl's conservative algorithms. In the case with neutrals, some power is lost to ionization and charge exchange interactions, indicated by the brown line. Since the input power is the same, the magnitude of losses to the endplates is slightly reduced. The sum of the source and losses for the neutrals case is given by the purple line, which also approaches zero, albeit with slightly worse errors. Presumably this is due to slightly larger fluctuations in the stored energy, arising from neutral interactions.

\begin{figure}
    \centering
    \includegraphics[width=.8\linewidth]{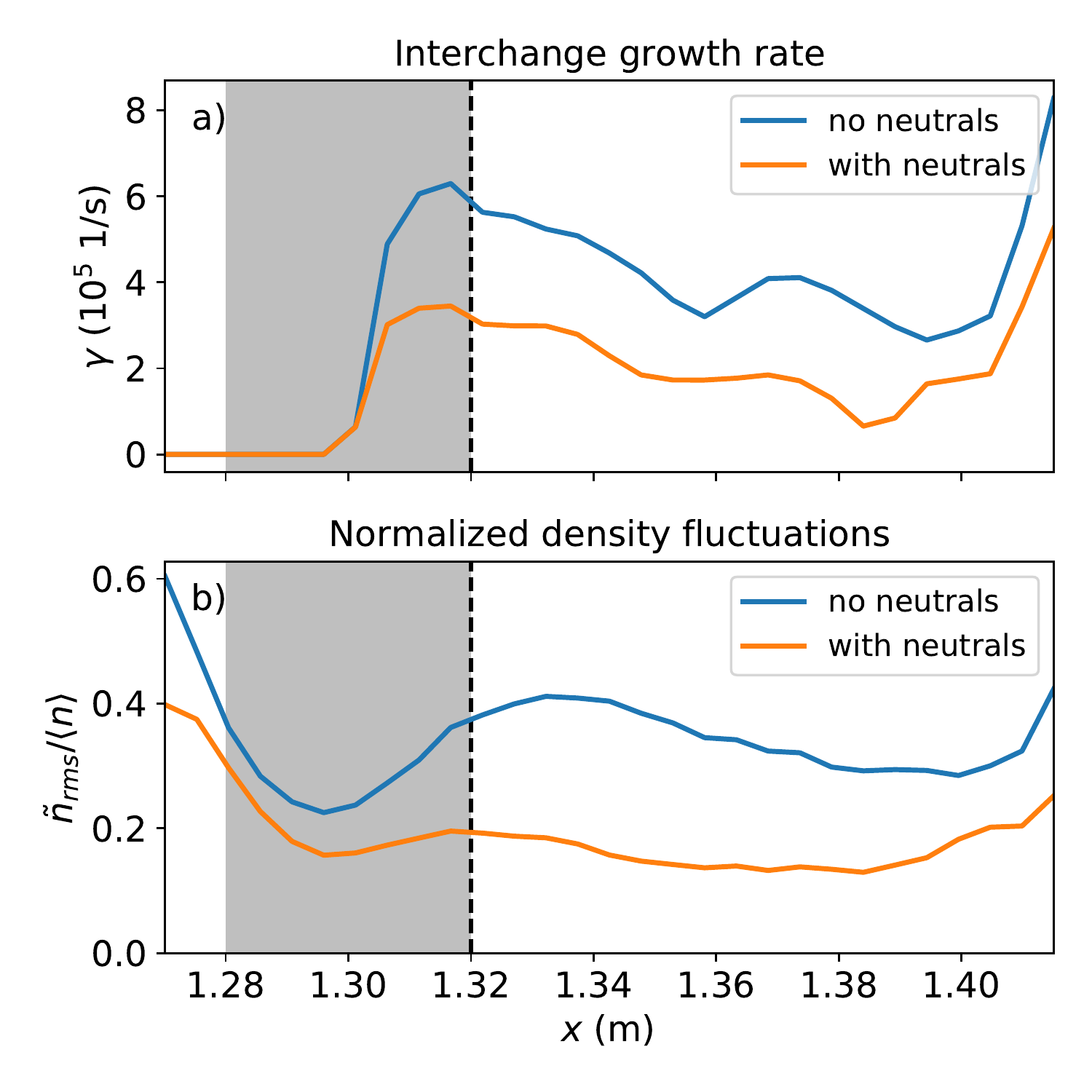}
    \caption{The linear interchange growth rates are calculated from the pressure profiles and compared in (a), where the simulation with neutrals shows reduced levels. The normalized density fluctuations are compared in (b) and are reduced in the simulation with neutrals.}
    \label{fig:gamI-dn}
\end{figure}

Since the turbulence is primarily driven by bad curvature, we have calculated the linear interchange growth rate $\gamma_I = c_s\sqrt{2/(L_p x)}$, where $c_s = \sqrt{(T_e + T_i)/m_i}$ is the local sound speed and $L_p = \max\left(0,-\left( \frac{1}{p} \frac{dp}{dx}\right)^{-1}\right)$ is the pressure gradient scale length. The growth rates are compared in Fig.~\ref{fig:gamI-dn}a, and the curvature drive is reduced in the case with neutrals due to the flattening of the pressure profile. There is a corresponding reduction in normalized rms density fluctuation values, shown in Fig.~\ref{fig:gamI-dn}b.

\begin{figure}
    \centering
    \includegraphics[width=.8\linewidth]{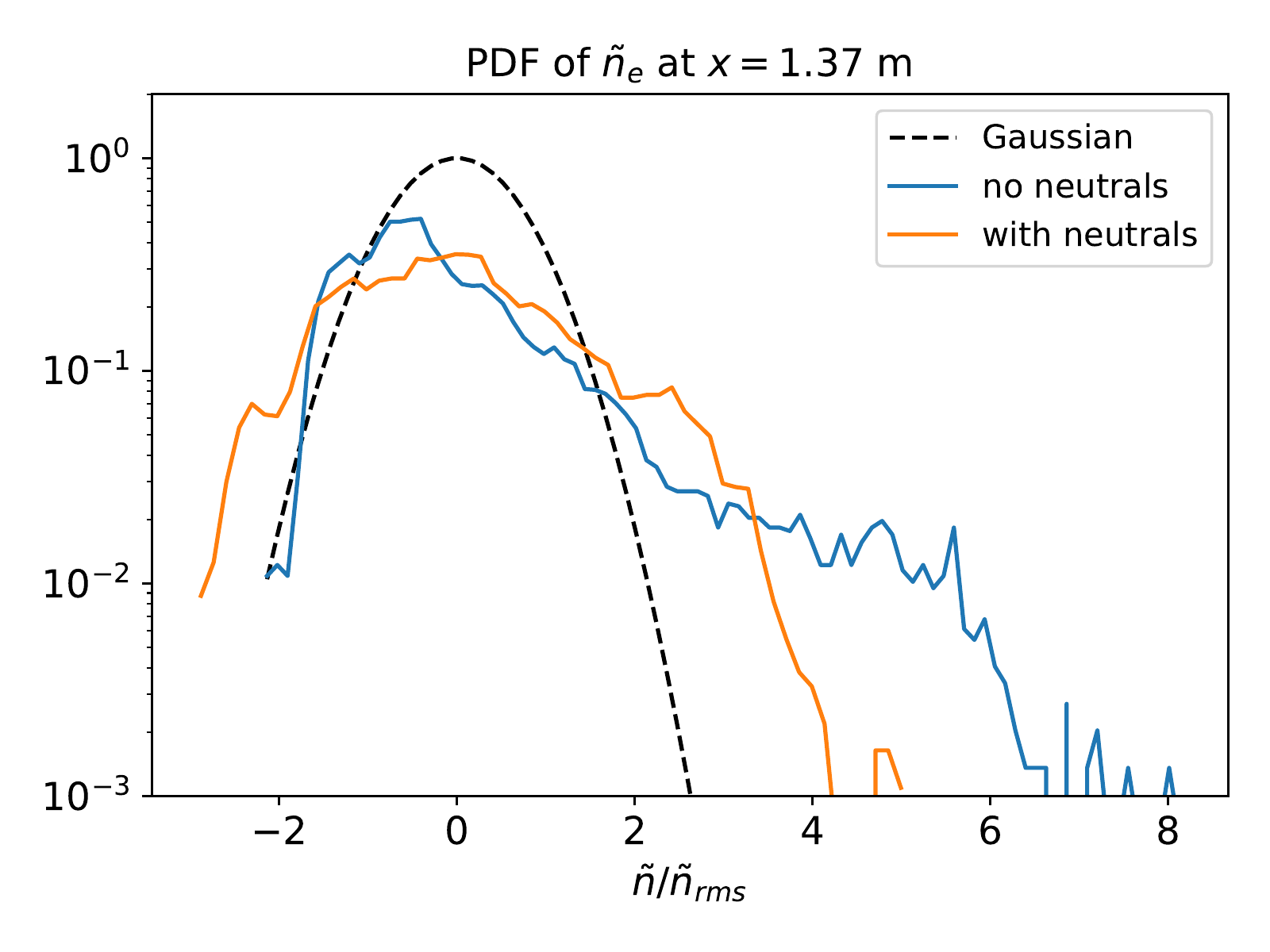}
    \caption{Probability distribution functions (PDFs) of electron density turbulence fluctuations at radial location $x=1.37$ m are compared to Gaussian distribution (dashed curve).}
    \label{fig:pdf-dn}
\end{figure}

\begin{figure}
    \centering
    \includegraphics[width=.8\linewidth]{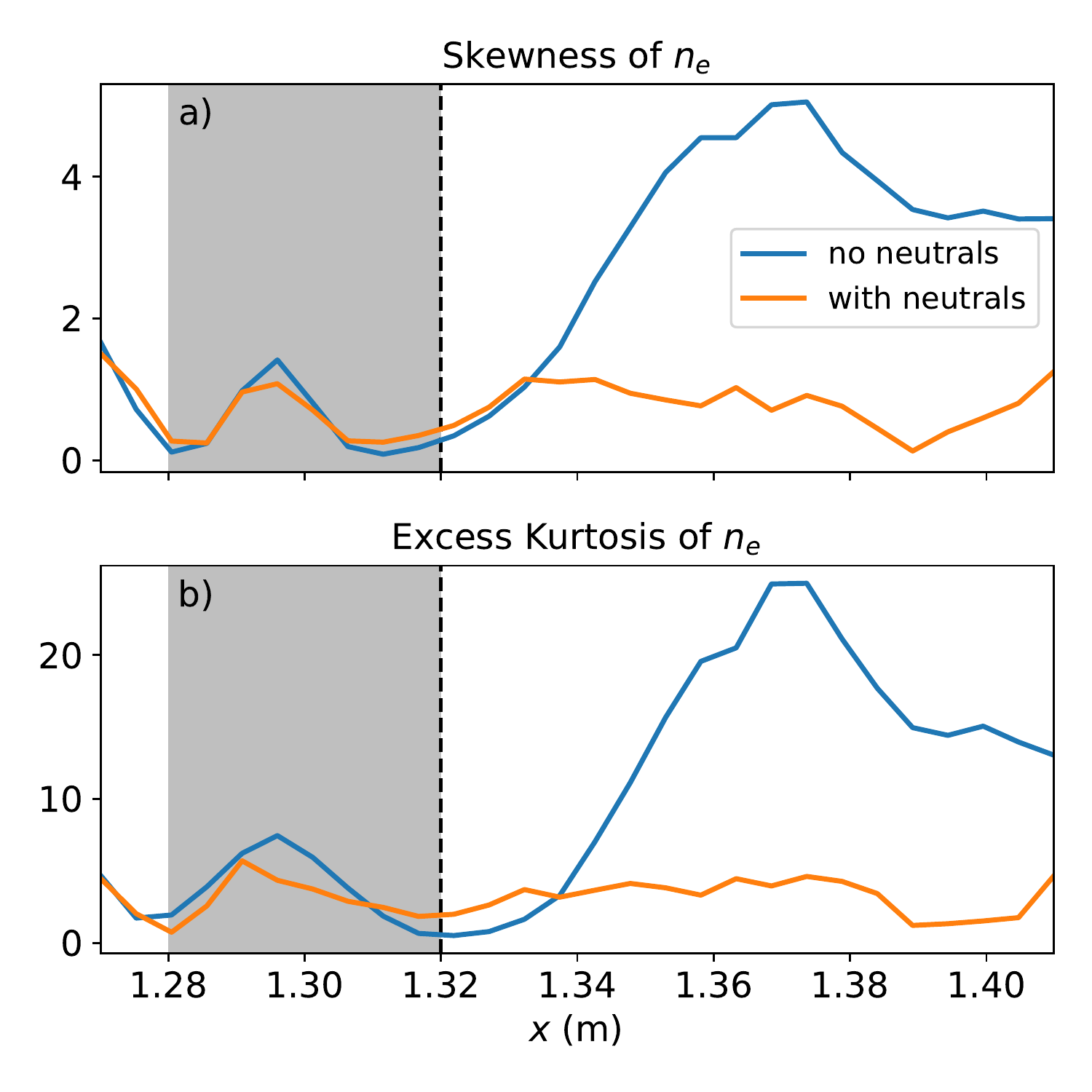}
    \caption{Skewness (a) and excess kurtosis (b) of electron density fluctuation PDFs are compared.}
    \label{fig:skew-kurt-dn}
\end{figure}

Turbulence statistics at the midplane, such as the probability distribution functions (PDFs) of turbulence fluctuations, have also been investigated. For the PDF, density fluctuations are measured over the time interval 200--400 $\mu$s at a particular radial value and normalized to the rms value of fluctuations ($\tilde{n}/\tilde{n}_{rms}$. Examples of PDFs calculated at $x=1.37$ m are shown in Fig.~\ref{fig:pdf-dn} and compared to a Gaussian distribution (dashed curve). The case without neutrals is more skewed in the positive direction and has a longer positive tail. Positive skewness and excess kurtosis of the density fluctuation PDFs are characteristics of blobby turbulence.\cite{dippolito2011} Skewness and excess kurtosis are calculated as $E[\tilde{n}^3]/\sigma^3$ and $E[\tilde{n}^4]/\sigma^4 - 3$, respectively, where $E[...]$ is the expectation value, given by the PDF. The skewness and excess kurtosis of density fluctuation PDFs are shown in Fig.~\ref{fig:skew-kurt-dn}a and Fig.~\ref{fig:skew-kurt-dn}b, respectively. The skewness and kurtosis for the case with neutrals are nearly constant in $x$. This is consistent with the weaker density gradients and lower fluctuations observed previously, as there are fewer turbulence events that deviate greatly from the mean. For example, in the plot of density fluctuation profiles (Fig.~\ref{fig:gamI-dn}b), the fluctuation levels from the case with neutrals are nearly constant in $x$ while they increase more noticeably at large radii for the case without neutrals case. We note that the increase in skewness and kurtosis at large radii in the case without neutrals is more consistent with experimental observations.\cite{dippolito2011}

\begin{figure}
    \centering
    \includegraphics[width=.8\linewidth]{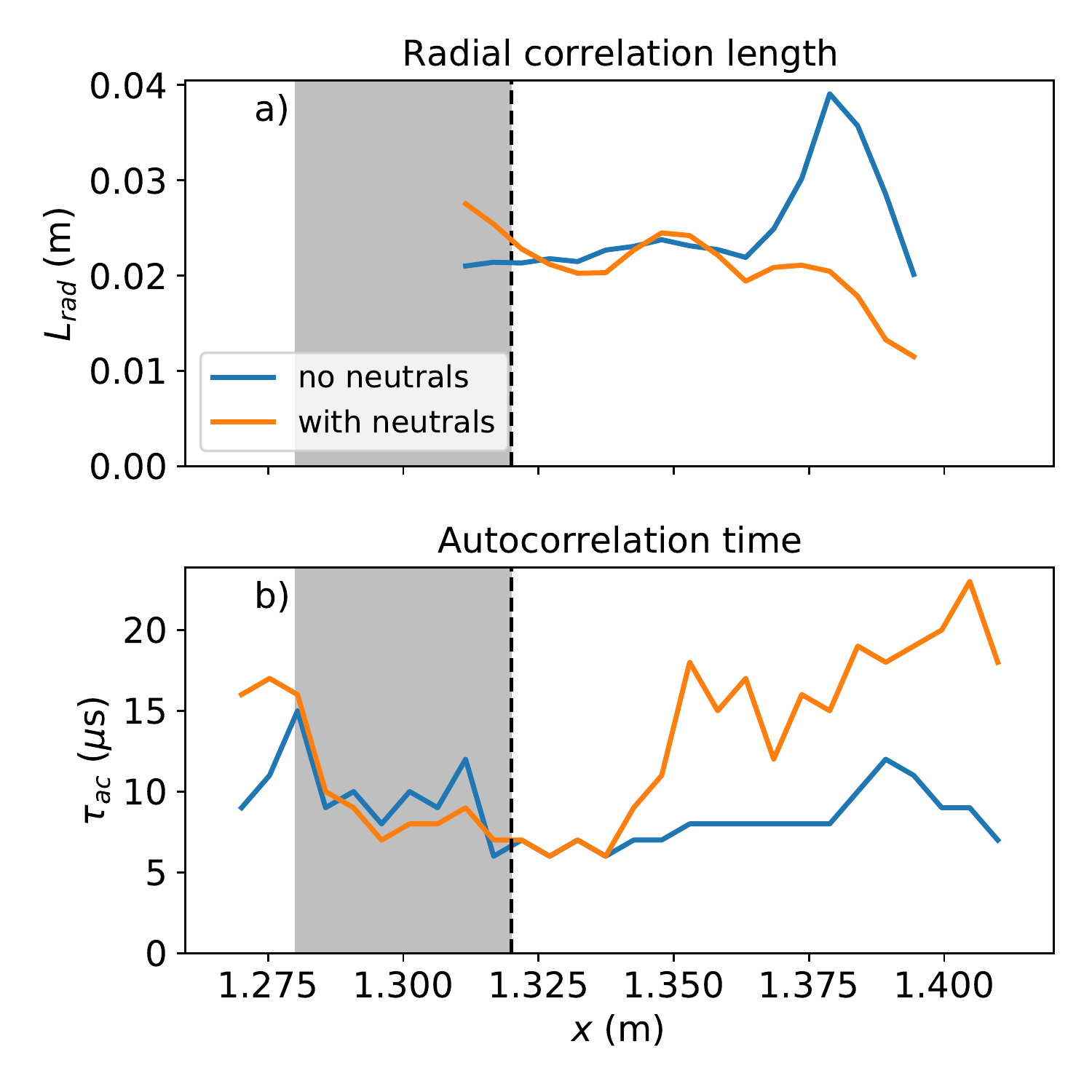}
    \caption{Comparisons of radial correlation lengths (a) and autocorrelation times (b).}
    \label{fig:ac-Lrad}
\end{figure}

Radial correlation lengths are compared in Fig.~\ref{fig:ac-Lrad}a. We find no change in the radial correlation lengths, except at large radii, where they decrease for the case with neutrals. This can be used as a proxy to determine blob size. Autocorrelation times, $\tau_{ac}$, are compared in Fig.~\ref{fig:ac-Lrad}b, which are calculated as the $e$-folding time of the autocorrelation functions. The autocorrelation function is calculated as $C(\tau) = \langle \tilde{n}(t)\tilde{n}(t + \tau)\rangle/\langle \tilde{n}(t)^2 \rangle$, where brackets denote a time average. Autocorrelation times increase in the simulation with neutrals, suggesting longer temporal coherency of turbulent structures. Both radial correlation lengths and autocorrelation times were calculated with data from the interval 200--400 $\mu$s.

The proof-of-concept simulation with neutrals incorporated a self-consistent source via ionization. Neutral interactions impacted the steady-state profiles and turbulence statistics in noticeable ways, demonstrating the importance of incorporating them in predictive modeling. Additional studies have being carried out in which an additional volumetric source to approximate the ionization source is added in simulations without the neutral species, in order to isolate sourcing effects from charge exchange collisions. These demonstrated that most of the differences in the comparison presented herein were due to sourcing. However, differences in properties of the blob dynamics remained. The full details of these results are outside the scope of this paper and will be presented in a future publication.

\section{Conclusions}\label{sec:concl}

We have presented the coupling of a continuum gyrokinetic plasma model to a continuum kinetic model for neutrals within \gkyl. A continuum code has the benefit of reduced noise and improved accuracy at a fixed resolution. Simplified models of electron impact ionization, charge exchange and wall recycling have been included. These models have been verified against analytic theory and show good agreement with DEGAS2 in benchmark tests. 

A proof-of-concept has been demonstrated in a high dimensional simulation of the NSTX SOL with simplified helical magnetic geometry. This simulation included gyrokinetic electron and ion species in three spatial dimensions and two velocity space dimensions coupled to Vlasov neutrals in three spatial dimensions and three velocity space dimensions. In a comparison to a baseline simulation without neutrals and a midplane source identical to the simulation with neutrals, the latter exhibited a flatter density profile, reduced turbulence fluctuations, and longer auto-correlation times.

The flatter density profile contributed to a flattening of the pressure profile and a corresponding reduction in the linear interchange growth rate. The normalized density fluctuations were also reduced. In Fig.~\ref{fig:skew-kurt-dn} skewness and kurtosis were reduced in the simulation with neutrals, also likely due to the increased density and the flatter density profile on the low-field side of the domain. The sharper density gradient in the case without neutrals means that high-density coherent turbulent structures that originate from the source region deviate more from the background density as they move radially outward, leading to a stronger positive skewness than in the simulation with neutrals. Large values of skewness and excess kurtosis are associated with greater intermittency, i.e. blobby turbulence.\cite{dippolito2011} The increased correlation times for the case with neutrals, seen in Fig.~\ref{fig:ac-Lrad}b, suggest turbulent structures are more coherent, which may be affecting the nature of the turbulence and the resulting radial transport of particles.. 



Collisionality and magnetic geometry also have important effects on SOL turbulence,\cite{Myra2006} and subsequent simulations will include these features more realistically. A reduced Spitzer collision frequency that was constant in time and space was used in these simulations to reduce computational costs. The collision frequency can be calculated dynamically in \gkyl\ based on local densities and temperatures, and this feature will be included in simulations with neutrals. Work is ongoing to extend the neutral model to be compatible with the general geometry features available in \gkyl, which are outlined in Chapter 5 of Ref.~\citenum{Mandell2021thesis}. Chapter 4 of the same reference also demonstrates the impact of electromagnetic effects on blob dynamics, and this could also be studied in simulations that include neutral interactions.

Furthermore, the effect of neutral interactions in high-recycling or detached SOL scenarios is important. Modeling these scenarios with \gkyl\ will require the ability to represent low neutral temperatures on the velocity space grid. A non-uniform grid is currently under development and will help with this issue. Models for recombination and radiation will eventually be included to model detached conditions. 
This will enable us to incorporate new and existing features such as increased collisionality, realistic geometry and electromagnetic plasma species in simulations with neutrals. 

Some of the benefits of Monte Carlo neutral codes are challenging to reproduce with a continuum code. MC codes can simulate many species simultaneously, including molecules and excited metastable states. They can also easily model short-lived species or those that only occupy part of the computational domain. On the other hand, continuum codes have the advantage of being able to model scenarios with a large density variation whose dynamical evolution significantly impacts physical outcomes. The \gkyl\ model can self-consistently explore the effect of neutral interactions on plasma profiles and turbulence statistics, while most MC codes couple to reduced models of plasma transport and do not consider effects on turbulence. With forthcoming GPU capabilities, \gkyl\ will be able to model plasma species, a neutral species, and an impurity species at reasonable computational cost, and results could be used to improve reduced models of plasma transport in MC codes.

\appendix

\section{Getting \gkyl\ and reproducing results} \label{apx:get-gkyl}
Readers may reproduce our results and also use Gkeyll for their applications. The code and input files used here are available online. Full installation instructions for Gkeyll are provided on the Gkeyll website.\cite{gkeyllWeb} The code can be installed on Unix-like operating systems (including Mac OS and Windows using the Windows Subsystem for Linux) either by installing the pre-built binaries using the conda package manager
(\hyperlink{https://www.anaconda.com}{https://www.anaconda.com}) or building the code via sources. The input files used here are under version control and can be obtained from the repository at \hyperlink{https://github.com/ammarhakim/gkyl-paper-inp/tree/master/2022_PoP_neutrals
}{https://github.com/ammarhakim/gkyl-paper-inp/tree/master/2022\_PoP\_neutrals}.

\section{Neutrals in simplified helical geometry} \label{apx:geo}

We now describe some subtleties of the coordinate system employed in \gkyl\ and show how the Vlasov neutral phase-space grid used in the NSTX simulations is consistent with various geometric assumptions. In this work and previous gyrokinetic simulations with the \gkyl\ code,\cite{Shi2017thesis,shi2019full,bernard2019,Bernard2020,mandell2020electromagnetic} we have used a simplified helical model of the magnetic field. The mapping from the cylindrical coordinates $(R, \varphi, Z)$ to those used in \gkyl\ $(x, y, z)$ is described in Refs.~\citenum{shi2019full,bernard2019} and Chapter 4 of Ref.~\citenum{Mandell2021thesis} as:
\begin{equation}
    x = R, \hspace{.5cm}
    z = \frac{Z}{\sin \vartheta} = \frac{L_c}{H}Z, \hspace{.5cm}
    y = {R_c}{\sin \vartheta}\left(\varphi - \frac{Z}{R_c}\cot \vartheta\right), \label{eq:helical-map}
\end{equation}
where $R_c$ is the center of the $x$-domain, $H$ is the vertical height, $L_c$ is the connection length, $\vartheta$ is the field line pitch angle, and $\sin \vartheta = B_v/B = H/L_c$. Nearly all geometrical factors arising from this non-orthogonal coordinate system are neglected in this simplified helical model, which effectively takes the $B_v \ll B$ limit. Only the assumption that perpendicular gradients of the distribution function are much stronger than parallel gradients is retained. By neglecting such geometric factors, simulations in the non-orthogonal helical field coordinate system defined by Eq.~\ref{eq:helical-map} can also be interpreted as occurring in a simpler orthogonal coordinate system with only a toroidal magnetic field, as given by the mapping: 
\begin{equation}
    x = R, \hspace{1cm}
    y = Z, \hspace{1cm}
    z = R_c \varphi. \label{eq:ortho-map}
\end{equation}
\begin{figure}
    \centering
    \includegraphics[width=.5\linewidth]{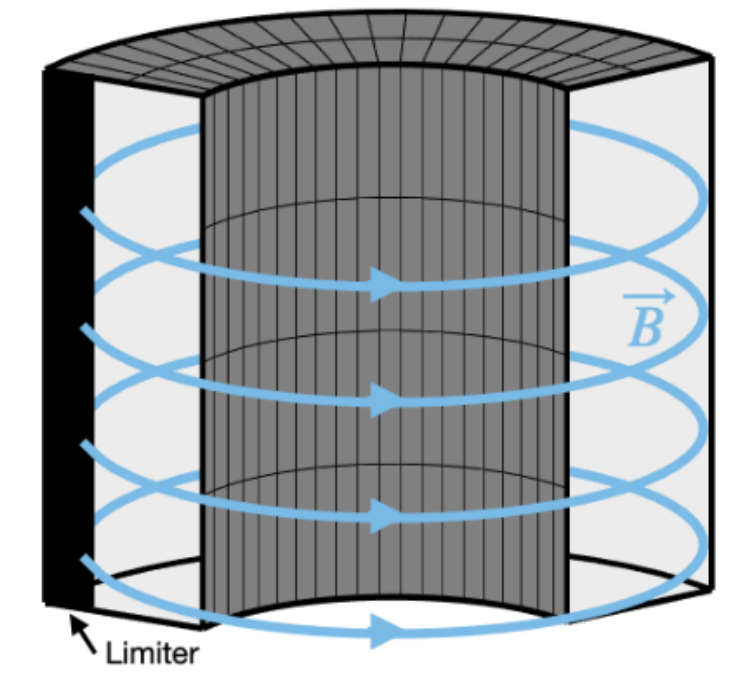}
    \caption{Diagram of the magnetic geometry corresponding to the geometry assumptions made in the simulations presented in Sec.~\ref{sec:nstx}. (Image from N. Mandell, PhD Thesis, Princeton University, 2021; licensed under a Creative Commons Attribution (CC BY) license.)}
    \label{fig:toroidal-geo}
\end{figure}
This is depicted in Fig.~\ref{fig:toroidal-geo} which originally appeared in Ref.~\citenum{Mandell2021thesis}. The relationship between these two geometry interpretations was pointed out in Appendices 5.A–5.B of Ref.~\citenum{Mandell2021thesis}. Note that both coordinate systems are field-aligned, since gradients parallel to the magnetic field are expressed in terms of derivatives involving only one coordinate, $\nabla_{||} = (1/|\vec{B}|) \vec{B} \cdot \nabla \propto \partial / \partial z$. The difference is that the direction of the magnetic field in physical space is slightly different in the two coordinate systems. In the geometry of Eq.~\ref{eq:ortho-map}, the field-line coordinate $z$ is purely toroidal and the binormal coordinate $y$ is in the vertical direction. The sheath conducting boundary conditions along the $z$ direction are then effectively being applied along a limiter that extends in the vertical direction. The simulated field lines start at a toroidally-localized limiter, but may wrap more (or less) than $2 \pi R$ around the torus before hitting a limiter again. The gradient basis vectors are
\begin{equation}
    \nabla x = \hat{\bm{R}}, \hspace{1cm}
    \nabla y = \hat{\bm{Z}}, \hspace{1cm}
    \nabla z = \frac{R_c}{x}\hat{\bm{\varphi}}, 
\end{equation}
The present method of coupling the Vlasov neutral solver to the gyrokinetic solver is consistent with this configuration. The neutral species share the same field-line-following coordinate system for configuration space $(x,y,z)$, and we assume an orthogonal cylindrical coordinate system for the velocity space, $(v_R, v_Z, v_\varphi)$. The streaming term in the Vlasov equation (\ref{eq:vlasov}) becomes
\begin{eqnarray}
    \bm{v} \cdot \nabla f &=& \bm{v} \cdot \left( \nabla x \pd{f}{x} + \nabla y \pd{f}{y} + \nabla z \pd{f}{z} \right), \nonumber \\
    &=& v_R \pd{f}{x} + v_Z \pd{f}{y} + v_\varphi \frac{R_c}{x} \pd{f}{z}, \nonumber \\
    &\approx& v_R \pd{f}{x} + v_Z \pd{f}{y} + v_\varphi \pd{f}{z}, 
\end{eqnarray}
where we have assumed $R_c/x \approx 1$ for a flux-tube-like domain localized around $x=R_c$. This configuration is sufficient for this paper in showing the overall feasibility of this approach and doing a first study of neutral-turbulence interaction with this code. A more realistic helical geometry with magnetic shear is available in \gkyl\ and has been described in Chapter 5 of Ref.~\citenum{Mandell2021thesis}. Eventually, the neutral model will be incorporated in simulations with more realistic geometry and the streaming term in the Vlasov equation will be updated to include additional geometric terms that arise from the associated gradient basis vectors.

\section{Derivation of 1X1V analytic solution for verification tests} \label{apx:theory}
To verify models of neutral interactions, we derive theoretical predictions of steady-state profiles. First consider the 1X1V neutral equation in the form
\begin{equation}
	\frac{\partial f}{\partial t} +v_z \frac{\partial f}{ \partial z} = -\nu \left(f(z,v_z)-g(z,v_z)\right), \label{eq:gen-neut-coll}
\end{equation}
where
	\begin{equation}
	 g(z,v_z)= \frac{f_i}{n_i} \int d v_z' f(z,v_z').
	\end{equation}
In the case with only ionization, $f_i = 0$. We seek a steady-state solution:
\begin{equation}
	v_z \frac{\partial f}{ \partial z} = -\nu \left(f(z,v_z)-g(z,v_z)\right). \label{eq:neut-coll-ss}
\end{equation}
Consider a finite, symmetric domain, $z = [-L/2,L/2]$, when solving Eq.~\ref{eq:neut-coll-ss}. We now have solutions for the left-hand and right-hand sides of the domain: 
\begin{equation}
    f(z,v_z)= \left\{
        \begin{array}{ll}
        C_+(z,v_z)e^{-\frac{\nu}{v_z}(z+L/2)} & v_z>0 \\
    C_-(z,v_z)e^{-\frac{\nu}{v_z}(z-L/2)} & v_z<0.
    \end{array} \right .
\end{equation}
This becomes
\begin{equation}\label{key}
	 C_+(z,v_z) =C_+(-L/2,v_z)+ \int_{-L/2}^z dz' \frac{\nu}{v_z} g(z',v_z)e^{\frac{\nu}{v_z}(z'+L/2)}
\end{equation}
\begin{equation}\label{key}
	 C_-(z,v_z) =C_-(L/2,v_z)- \int_z^{L/2} dz' \frac{\nu}{v_z} g(z',v_z)e^{\frac{\nu}{v_z}(z'-L/2)},
\end{equation}
which gives
\begin{widetext}
\begin{equation}
    f(z,v_z)=\left\{
        \begin{array}{ll}\left( C_+(-L/2,v_z)+ \int_{-L/2}^z dz' \frac{\nu}{v_z} g(z',v_z)e^{\frac{\nu}{v_z}(z'+L/2)}\right)e^{-\frac{\nu}{v_z}z} & v_z>0 \\
    \left(C_-(L/2,v_z)- \int_z^{L/2} dz' \frac{\nu}{v_z} g(z',v_z)e^{\frac{\nu}{v_z}(z'-L/2)}\right)e^{-\frac{\nu}{v_z}(z-L)} & v_z<0. 
    \end{array} \right .
\end{equation}
To simplify notation, we introduce $f(-L/2,v_z>0)=f^{BC}_{-L/2}(v_z)$, and $f(L/2,v_z<0)=f^{BC}_{L/2}(v_z)$ which are the distributions numerically imposed at the boundary conditions, such that
\begin{equation}
    f(z,v_z)=\left\{
        \begin{array}{ll}\left(f^{BC}_{-L/2}(v_z)+ \int_{-L/2}^z dz' \frac{\nu}{v_z} g(z',v_z)e^{\frac{\nu}{v_z}(z'+L/2)}\right)e^{-\frac{\nu}{v_z}z} & v_z>0 \\
    \left(f^{BC}_{L/2}(v_z) - \int_z^{L/2} dz' \frac{\nu}{v_z} g(z',v_z)e^{\frac{\nu}{v_z}(z'-L/2)}\right)e^{-\frac{\nu}{v_z}(z-L)} & v_z<0
    \end{array} \right .
\end{equation}
The density is thus given by
\begin{equation}
    n(z)= n^{BC}_{-L/2}(z) +\int_{-L/2}^z dz'n(z')\beta^+_i(z,z')+ n^{BC}_{L/2}(z)-\int_z^{L/2} dz'n(z')\beta^-_i(z,z'), \label{eq:sym-density}
\end{equation}
where 
\begin{eqnarray}
n^{BC}_{-L/2}(z)&=&\int_0^{\infty} d v_z' \, f^{BC}_{-L/2}(v_z') e^{-\frac{\nu}{v_z'}(z+L/2)}\\
n^{BC}_{L/2}(z)&=&\int_{-\infty}^0 d v_z' \, f^{BC}_{L/2}(v_z') e^{-\frac{\nu}{v_z'}(z-L/2)}\\
\beta^+_i(z',z)&=&\int_0^\infty d v_z' \, \frac{\nu}{v_z'} G_i(z',v_z')  e^{\frac{\nu}{v_z'}(z'-z)}\\
\beta^-_i(z',z)&=&\int_{-\infty}^0 d v_z' \, \frac{\nu}{v_z'} G_i(z',v_z')  e^{\frac{\nu}{v_z'}(z'-z)},
\end{eqnarray}
and $G_i(z,v_z) = f_i(z,v_z)/n_i(z)$.
\end{widetext}

\begin{acknowledgments}
T. Bernard acknowledges helpful discussions with J. Boedo and F. Scotti. 

This material is based upon work supported by the U.S. Department of Energy, Office of Science, Office of Fusion Energy Sciences, Theory Program, under Award No.~DE-FG02-95ER54309. J. Juno acknowledges support from a NSF Atmospheric and Geospace Science Postdoctoral Fellowship (Grant No. AGS-2019828). J. Guterl acknowledges DOE grant DE-SC0018423. Computing resources were provided by the Extreme Science and Engineering Discovery Environment (XSEDE), which is supported by National Science Foundation grant number ACI-1548562 and also by the National Energy Research Scientific Computing Center (NERSC), which is an Office of Science User Facility supported under Contract DE-AC02-05CH11231. 

-- Disclaimer -- This report was prepared as an account of work sponsored by an agency of the United States Government. Neither the United States Government nor any agency thereof, nor any of their employees, makes any warranty, express or implied, or assumes any legal liability or responsibility for the accuracy, completeness, or usefulness of any information, apparatus, product, or process disclosed, or represents that its use would not infringe privately owned rights. Reference herein to any specific commercial product, process, or service by trade name, trademark, manufacturer, or otherwise, does not necessarily constitute or imply its endorsement, recommendation, or favoring by the United States Government or any agency thereof. The views and opinions of authors expressed herein do not necessarily state or reflect those of the United States Government or any agency thereof.
\end{acknowledgments}

\section*{Data Availability Statement}

The data that support the findings of this study are available from the corresponding author upon reasonable request.

\end{document}